\documentclass[sigconf, 10pt]{acmart}
\usepackage{prp-macros}
\usepackage{booktabs}
\usepackage{subcaption}
\usepackage{url}
\usepackage{multicol}
\usepackage{multirow}
\usepackage{enumitem}
\usepackage{balance}
\usepackage{hhline}
\usepackage{listings}
\usepackage{enumitem}
\usepackage{xspace}
\usepackage{pifont}
\usepackage{framed}
\usepackage{algorithm}
\usepackage[noend]{algpseudocode}
\usepackage[skip=1ex]{caption}
\usepackage{cancel}
\usepackage{xcolor}
\usepackage{colortbl}
\usepackage{array}

\definecolor{light-gray}{gray}{0.80}
\newcolumntype{M}[1]{>{\centering\arraybackslash}m{#1}}
\setlist{leftmargin=5.5mm}
\lstnewenvironment{verblisting}{
	\lstset{
		basicstyle=\footnotesize\ttfamily,
		tabsize=2,
		breaklines=true
	}
}{}

\lstnewenvironment{codelisting}{
	\lstset{
		language=[Sharp]C,
		basicstyle=\footnotesize\ttfamily\linespread{0.8},
		tabsize=2,
		keywordstyle=\color{blue}\ttfamily,
		stringstyle=\color{red}\ttfamily,
		commentstyle=\color{magenta}\ttfamily,
		morecomment=[l][\color{magenta}]{\#},
		breaklines=true
	}
}{}

\newcommand{\eat}[1]{}

\makeatletter
\newcommand{\removelatexerror}{\let\@latex@error\@gobble}
\makeatother

\newtheorem{theorem}{Theorem}

\newtheorem{definition}{Definition}

\usepackage{lipsum}
\usepackage{mdframed}
\usepackage{tikz}
\usetikzlibrary{fit,calc}
\newcommand{\sys}{\textsc{Woodblock}\xspace}

\newfont{\ttlfnt}{phvb at 18pt}
\definecolor{light-gray}{gray}{0.95}
\definecolor{mid-gray}{gray}{0.85}
\definecolor{green}{RGB}{0,176,80}
\definecolor{darkred}{rgb}{0.7,0.25,0.25}
\definecolor{darkgreen}{rgb}{0.15,0.55,0.15}
\definecolor{darkblue}{rgb}{0.1,0.1,0.5}
\definecolor{orange}{RGB}{237,125,49}
\definecolor{blue}{RGB}{68,114,196}
\definecolor{pop}{RGB}{0,21,245}
\definecolor{realblue}{RGB}{0,0,255}

\usepackage{hyperref}       % hyperlinks
\hypersetup{
    hidelinks=false,
    colorlinks=true,
    filecolor=magenta,
    linkcolor=darkred,
    citecolor=darkred,
    urlcolor=darkred,
}

\newcommand{\revision}[1]{{#1}}
\newcommand{\removedfromrevision}[1]{}
\newcommand{\removedfromcamready}[1]{}

\definecolor{ReviewerDarkGray}{HTML}{37474F}
{\endlist\end{mdframed}}

\usepackage{ifthen}
\newcommand{\DONE}[1]{}

\newcommand{\ignore}[1]{}

\newcounter{programlinenumber}

\newboolean{TR}
\setboolean{TR}{false}
\ifthenelse{\boolean{TR}}{

\newcommand{\TrOnly}[1]{#1}
\newcommand{\SubOnly}[1]{}
\newcommand{\TrOnlyInFootnote}[1]{#1}
\newcommand{\TrOnlyInTable}[1]{#1}}
{

\newcommand{\TrOnly}[1]{}
\newcommand{\SubOnly}[1]{#1}
\newcommand{\TrOnlyInFootnote}[1]{}
\newcommand{\TrOnlyInTable}[1]{}}

\graphicspath{{images/}}
\usepackage{wrapfig}
\usepackage{xspace}

\makeatletter
\def\blfootnote{\xdef\@thefnmark{}\@footnotetext}
\makeatother

\newtheorem{prb}{Problem}
\newcommand{\qdrt}{\textsf{qd-tree}\xspace}
\newcommand{\qdrts}{\textsf{qd-trees}\xspace}
\newcommand{\qdt}{\textsf{Qd-tree}\xspace}
\begin{document}

\setcopyright{acmcopyright}

\copyrightyear{2020}
\acmYear{2020}
\setcopyright{acmcopyright}
\acmConference[SIGMOD'20]{Proceedings of the 2020 ACM SIGMOD International Conference on Management of Data}{June 14--19, 2020}{Portland, OR, USA}
\acmBooktitle{Proceedings of the 2020 ACM SIGMOD International Conference on Management of Data (SIGMOD'20), June 14--19, 2020, Portland, OR, USA}
\acmPrice{15.00}
\acmDOI{10.1145/3318464.3389770}
\acmISBN{978-1-4503-6735-6/20/06}
% \copyrightyear{2018}
% \acmYear{2018}
% \acmDOI{10.1145/1122445.1122456}

% %% These commands are for a PROCEEDINGS abstract or paper.
% \acmConference[Woodstock '18]{Woodstock '18: ACM Symposium on Neural
%   Gaze Detection}{June 03--05, 2018}{Woodstock, NY}
% \acmBooktitle{Woodstock '18: ACM Symposium on Neural Gaze Detection,
%   June 03--05, 2018, Woodstock, NY}
% \acmPrice{15.00}
% \acmISBN{978-1-4503-9999-9/18/06}

\settopmatter{printacmref=true}
\fancyhead{}

\title{Qd-tree: Learning Data Layouts for Big Data Analytics}
\author{Zongheng Yang$^{\S\ast}$, Badrish Chandramouli, Chi Wang, Johannes Gehrke$^\dagger$, Yinan Li,}
\author{Umar Farooq Minhas, Per-\AA{ke} Larson, Donald Kossmann, Rajeev Acharya$^\dagger$}

\gdef\authors{Zongheng Yang, Badrish Chandramouli, Chi Wang, Johannes \linebreak Gehrke, Yinan Li, Umar Farooq Minhas, Per-\AA{ke} Larson, Donald Kossmann, Rajeev Acharya}

\affiliation{%
  \institution{Microsoft Research \qquad\qquad $^\dagger$Microsoft \qquad\qquad $^\S$University of California, Berkeley}
}

\email{zongheng@cs.berkeley.edu,  {badrishc, chiw, johannes, yinali}@microsoft.com,}
\email{{ufminhas, a-palars, donaldk, rajeevac}@microsoft.com}

% \title{Learning Data Layouts for Large Analytical Workloads}

\begin{abstract}
% With the increasing importance of big data analytics, the need to run queries very quickly
% on large datasets is becoming increasingly important.
Corporations today collect data at an unprecedented and accelerating scale, making
the need to run queries on large datasets increasingly important.
Technologies such as columnar block-based data organization and compression have
become standard practice in most commercial database systems. % today.
However, the problem of best assigning records to data blocks on storage is
still open.
For example, today's systems usually partition data by arrival time into row
groups, or range/hash partition the data based on selected fields. For a given
workload, however, such techniques are unable to optimize for the important
metric of the \emph{number of blocks accessed} by a query. This metric directly
relates to the I/O cost, and therefore performance, of most analytical
queries. Further, they are unable to exploit additional available storage to
drive this metric down further.

In this paper, we propose a new framework called a \emph{query-data routing tree}, or {\em\qdrt}, to address this problem, and propose two algorithms for their construction based on greedy and deep reinforcement learning techniques.
Experiments over benchmark and real workloads show that a \qdrt can provide physical speedups of more than an order of magnitude compared to current blocking schemes, and can reach within $2\times$ of the lower bound for data skipping based on selectivity, while providing complete semantic descriptions of created blocks.
% Experiments over benchmarks and real workloads indicate that we are able to outperform the quality of current blocking schemes by up to an order-of-magnitude, and reach within $2\times$ of the lower bound for data skipping based on selectivity, while providing complete semantic descriptions of created blocks.

\end{abstract}

%\begin{CCSXML}
%	<ccs2012>
%		<concept>
%			<concept_id>10002951.10002952.10003190.10003193.10003429</concept_id>
%			<concept_desc>Information systems~Database recovery</concept_desc>
%			<concept_significance>500</concept_significance>
%		</concept>
%		<concept>
%			<concept_id>10002951.10002952.10003190.10003195.10010836</concept_id>
%			<concept_desc>Information systems~Key-value stores</concept_desc>
%			<concept_significance>500</concept_significance>
%		</concept>
%	</ccs2012>	
%\end{CCSXML}
%\ccsdesc[500]{Information systems~Database recovery}
%\ccsdesc[500]{Information systems~Key-value stores}

\maketitle

\blfootnote{$^\ast$Research done during internship at Microsoft Research.}

\section{Introduction}
\label{sec:intro}
% The column width is: \the\columnwidth
The last decade has seen a huge surge in the volume of data collected for big data analytics. This trend has in turn driven up interest in building high-performance analytics systems that can answer queries over terabytes of data in seconds.

The first generation of analytics systems heavily leveraged fine-grained indexes such as B-Trees to accelerate query processing. However, more recently, the trend has shifted to using scan-oriented data processing strategies that exploit the high sequential bandwidth of modern storage devices. In an effort to reduce the amount of data read from disk, today's analytics systems typically split up the data into chunks or \emph{data blocks} in main memory or secondary storage. Further, they build an in-memory \emph{min-max index} that stores the minimum and maximum values per block, per field, and use it to only retrieve blocks relevant to a query. A retrieved block is fully scanned, and other blocks are entirely skipped.

% While we have made tremendous progress
While tremendous progress has been made
in organizing the data carefully within each block, with techniques such as columnar organization and compression, the problem of best assigning records to data blocks is relatively less explored. For example, today's production systems usually partition data by arrival time into  blocks called \emph{row groups}, or use \emph{hash} or \emph{range} partitioning of data based on selected fields. Such techniques are unable to optimize for the important metric of \emph{number of blocks accessed} (or rows touched, in case of variable sized blocks) by a given query workload. This metric directly relates to the I/O cost, and therefore performance, of the workload.

% We would like to create blocks
We desire blocks
with two properties: (1) a \emph{semantic description}, a precise description of its contents as a predicate over table fields; and (2) \emph{completeness}, a guarantee that a block contains \emph{all} tuples matching the given description. These properties are desirable for accelerating block retrieval and for using blocks as a local cache or partial view over remote data. Further, they allow us to consider advanced data layouts where a row is stored in more than one block, thereby exploiting additional storage budget (Sec.~\ref{sec:opt}). Such data redundancy can significantly reduce the number of blocks accessed by a query, but only if we have the completeness property for blocks. Without the completeness guarantee, all blocks overlapping with a query would need to be scanned, even if a single block contains all the tuples needed for that query. Recent research on row-grouping~\cite{bottom_up} proposes the use of data mining and clustering techniques to create data blocks; the blocks have semantic descriptions based on a feature bitmap vector, but the resulting blocks lack completeness (Sec.~\ref{sec:prelims} has more details).

\subsection{Our Solution}

\begin{figure}[t!]
\centering
\includegraphics[width=.85\columnwidth]{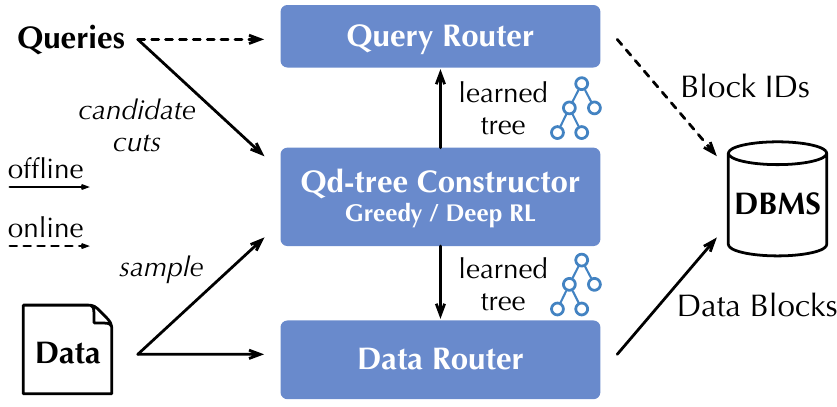}
% \vspace{-.05in}
\caption{\small System Architecture.\label{fig:arch}}
\end{figure}

In this paper, we propose a new framework to address the problem of data organization and query routing (Sec.~\ref{sec:prelims} and~\ref{sec:topdown}). Fig.~\ref{fig:arch} depicts our overall system architecture, \revision{targeting a traditional data warehouse scenario where data is stored on disk}. At the heart of the system is a data structure we call a \emph{query-data routing tree} (or \emph{\qdrt}). Briefly, a \qdrt is a binary tree where each node corresponds to some sub-space of the entire high-dimensional data space. The root of the tree corresponds to the entire data space. We perform a \emph{cut} of a node's data space in order to create its children.

 We can use a \qdrt for both block generation and query processing. Given a \qdrt and a dataset, we can route each tuple in the dataset through the \qdrt to assign them to data blocks of a minimum size (large blocks may be physically stored as multiple segments on storage). In other words, the leaves of the \qdrt correspond to data blocks. Each leaf has a semantic description, a predicate $p$ based on a conjunction of the cuts made to the data space as we traverse the routing tree from root to leaf. Data blocks created in this manner are also complete: a data block can be described as ``all tuples that match predicate $p$''. Further, given a set of \qdrt based blocks and an incoming query, we can use the \qdrt in conjunction with traditional min-max indexes to quickly locate and scan all blocks relevant to the query.

 Constructing the optimal \qdrt for a given dataset and workload is a hard problem. Our first approach (Sec.~\ref{sec:greedy}) uses a new greedy heuristic where, starting from the root, each cut is made based on locally available information. This provides routing trees of high quality with new approximation guarantees, but is unable to fully exploit long-term knowledge of tree quality. \revision{At the other extreme is \emph{dynamic programming} (\emph{DP}) or memoized search, which can find the optimal solution, but is infeasible given our large search space. Instead}, by exploiting deep reinforcement learning (\revision{RL}) for their construction, we show (Sec.~\ref{sec:rl}) that we can explore a larger search space and exploit any implicit lower dimensionality of data during \qdrt construction, thereby producing routing trees that significantly outperform the state-of-the-art, while still providing complete semantic descriptions for blocks. \revision{Our RL solution can be regarded as an approximate and accelerated memoized search method~\cite{powell2007approximate, powell2016perspectives},
   leading to higher efficiency than DP but with optimality close to DP.
   % leading to both efficiency and optimality between those of greedy and DP.
 } Deep RL-based \qdrt also forms a general framework: we show in Sec.~\ref{sec:opt} that it can easily be extended with newer types of cuts, data overlap, and data replication for even better block skipping.

 We perform a detailed evaluation (Sec.~\ref{sec:eval}) on a standard benchmark workload (TPC-H) and two real workloads. Our experiments show that \qdrt-based data layouts can yield physical speedups, compared to current block construction schemes, of up to $14\times$ per workload (or up to $180\times$ per query), and reach within $2\times$ of the lower bound for data skipping based on selectivity (the optimal solution, being a hard problem, is unknown). We cover related work in Sec.~\ref{sec:related} and conclude the paper in Sec.~\ref{sec:conclude}.

\subsection{Contributions}

To summarize, we make the following contributions:
\begin{itemize}[leftmargin=*]
    \item We introduce the \qdrt data structure and show how it can be used for data partitioning.
    \item We propose a greedy construction scheme for \qdrt, which builds the tree top-down starting from the root, and offers theoretical approximation guarantees.
    \item To overcome the limitations of the greedy approach, we introduce \sys, a deep reinforcement learning algorithm that learns to construct high-quality \qdrts.\removedfromrevision{ It  minimizes the number of blocks accessed for a given data and workload pair.}
    \item We discuss the generality of our tree and learning framework via potential extensions that are enabled by a \qdrt's semantic description and completeness properties.
    \item Through a detailed evaluation on benchmarks and real workloads, we show that a learned \qdrt exhibits excellent data skipping properties.

\end{itemize}

% Interestingly, one can view a \qdrt as bringing the power of \emph{learned multi-dimensional indexing} to modern block-based data organization.
% It allows us to express non-trivial block assignment strategies and locate relevant blocks quickly during query processing.
% The layout within each block is orthogonal to our work, affecting only the cost function; for instance, we can support columnar scan-optimized layouts as well as row-oriented layouts (possibly indexed) in each block.

Finally, note that one can view a \qdrt as a powerful workload-guided index for modern block-based big data analytics. It can express non-trivial block assignment strategies and can locate relevant blocks quickly during query processing. The layout within each block is orthogonal to the \qdrt itself, affecting only its cost function. For instance, a \qdrt can support columnar scan-optimized layouts as well as row-oriented layouts (possibly indexed) in each block.

\section{Preliminaries}
\label{sec:prelims}

\subsection{Problem Definition}
Given a set of tuples $V$, we aim to partition them into multiple blocks, such that the number of tuples required to scan for a workload is minimized, or equivalently, the number of tuples that can be skipped for the workload is maximized. Consider a partitioning $\mathcal{P}=\{P_1,\dots,P_k\}$ over $V$, i.e., $\mathcal{P}$ is a set of disjoint subsets of $V$ whose union is $V$. Each subset $P_i$ is called a block. We use $C(P_i)$ to denote the number of tuples that can be skipped when we execute all the queries in a workload $W=\{q_1,\dots,q_m\}$. For a scan-oriented system, it can be calculated as: %the following:
\begin{equation}\label{eq:c-func}
    C(P_i) = |P_i|\sum_{q\in W}S(P_i,q)
\end{equation}
where $S(P,q)$ is a binary function indicating whether partition $P$ can be skipped when processing query $q$. The definition of $S$ depends on the type of meta information maintained at each block. The most common type of meta information is the max-min filters, i.e., the maximal and minimal values of each dimension over all the tuples in a block. For this case, $S(P,q)=1$ if the hypercube defined by the max-min filters intersects with the range of query $q$. 

Given a workload $W$, the overall effectiveness of a partitioning $\mathcal{P}$ is measured by the total number of tuples skipped $C(\mathcal{P})=\sum_{P_i\in \mathcal{P}}C(P_i)$. 
Without constraints on block size, the number can be trivially maximized by putting every tuple in an individual block. For reasons like I/O batching and columnar compression, a real system requires blocks to have certain minimal size, e.g., 1 million tuples in SQL Server~\cite{DBLP:conf/sigmod/LarsonCFHMNPPRRS13}. We use $b$ to refer to this minimal size.

% The partitioning problem can be formulated as the following.
The partitioning problem is formulated as follows.
\begin{prb}[MaxSkip Partitioning]\label{def:maxskip}
Given a set $V$ of tuples, a workload $W$ of queries, a skipping function $S$, and a minimal block size $b$, find a partitioning $\mathcal{P}$ to maximize $C(\mathcal{P})$, s.t. $|P_i|\ge b$ for all $P_i\in \mathcal{P}$. 
\end{prb}

% \subsubsection{Dynamic vs. Static}
This formulation is appropriate for static data. To handle dynamically ingested data, it is desirable to learn a partitioning function from offline data, and apply the function to online data ingestion to save data reshuffling cost. %, such that the data reshuffling cost is saved.
\begin{prb}[Learned MaxSkip Partitioning]\label{def:learned}
Given a set $V$ of tuples, a workload $W$ of queries, a skipping function $S$, and a minimal block size $b$, find a partitioning function $F$, such that for the next $V$ tuples ingested, the partitioning $\mathcal{P}$ generated by $F(V)$ maximizes $C(\mathcal{P})$.
\end{prb}
In general, no partitioning function is guaranteed to work for future unseen data. In this work, we focus on the scenario where the current $V$ tuples have the same distribution as the next $V$ tuples. Therefore, solving Problem~\ref{def:learned} is reduced to solving Problem~\ref{def:maxskip} in addition with a descriptive partitioning function, such that any new tuple can be mapped to a right partition identifier. For efficiently ingesting data, we also desire the partitioning function to be lightweight to compute.

\subsection{Current Approaches}

\subsubsection{Date Partitioning}
In this basic partitioning scheme, we partition data by time of ingestion. The skipping function $S(P,q)=1$ if query $q$'s date range intersects with partition $P$, and is $0$ otherwise.

\subsubsection{Bottom-up Row Grouping}
This technique was proposed by Sun et al.~\cite{bottom_up}, and uses feature-based data skipping. Basically, each feature $f_i$ is a predicate over the data. $M$ features are extracted from the workload in the beginning using frequent pattern mining. Each block has a bitmap of length $M$, indicating whether predicate $f_i,i\in[M]$ is satisfied by any tuple in this block. If the $i$-th bit for this block is 0, i.e., no tuple satisfies $f_i$, then we can skip all queries subsumed by (i.e., stricter than) $f_i$. Sun et al.'s problem formulation is slightly different, requiring each partition to have equal size. They name the problem Balanced MaxSkip Partitioning, and prove its NP-hardness by reduction from hypergraph bisection. Using the same reduction technique, we can prove that Problem~\ref{def:maxskip} is NP-hard.

Sun et al.'s solution uses bottom-up clustering and is actually a solution to Problem~\ref{def:maxskip}, rather than the Balanced MaxSkip Partitioning problem. This is because the output of that algorithm has varying block sizes, and the sizes are no smaller than $b$\removedfromrevision{\footnote{This inconsistency exists in the original paper.}}. The algorithm converts tuples into unique binary feature vectors, and record the weight of each unique feature vector (row weight), as well as the number of queries subsumed by each feature (column weight). Initially every unique feature vector is in its own block. Then blocks are merged greedily using a heuristic criterion: in each iteration, a heuristic penalty is calculated for all pairs of blocks; and the pair with lowest penalty is chosen to be merged into a new block. Once the size of a block reaches $b$, it does not further merge with other blocks. Hence, merging eventually stops with every block having size no smaller than $b$.

This solution is shown to be more effective than date partitioning and simple multi-dimensional range partitioning.
There are several drawbacks of that approach. First, the heuristic penalty criterion used in the greedy algorithm only matches the optimization objective when the query sets subsumed by all features are disjoint. In general that assumption is not true. So choosing the pair of blocks minimizing the penalty does not necessarily maximize $C(\mathcal{P})$. Second, no theoretical guarantee is provided by the greedy merging algorithm. Third, the complexity of the algorithm is quadratic to the number of unique feature vectors, which can be as large as the number of tuples and grows exponentially with respect to the number of features. Practical application of this algorithm requires using a small number of features, which poses an additional challenge of selecting a good, small set of features. Last, while each block can be described using the ``OR'' of all the feature bitmap vectors contained in that block, such description is not complete. For example, there can be two blocks with identical bitmap description, and a new tuple does not have a deterministic destination partition using this description.

% LocalWords:  MaxSkip hypergraph al's

\section{\qdt}
\label{sec:topdown}

\def \treeA {\tikz[inner sep=2pt, 
% baseline=(T3.north)
] {
                  \node (b1) at (0,0) {$B_1$};
                  \node (b2) at (1,0) {$B_2$};
                  
                  \node (b3) at (2,0) {$B_3$};
                  \node (b4) at (3,0) {$B_4$};
                  
                  \node (n1) at (0.5,0.7) {{\sffamily mem=10GB?}};
                  \node (n2) at (2.5,0.7) {{\sffamily cpu<5\%?}};
                  
                  \node (n0) at (1.5,1.4) {{\sffamily cpu<10\%?}};

                  \draw (b1)--(n1)--(b2);
                  \draw (b3)--(n2)--(b4);
                  \draw (n1)--(n0)--(n2);
                  }
                  }
\begin{figure}[t]
        \begin{tikzpicture}[inner sep=3pt,text centered]
          \node (orig) {\treeA};
        \end{tikzpicture}
      \caption{An example \qdrt with four leaf blocks.\label{fig:tree-example}}
\end{figure}
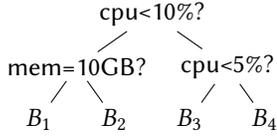

A \qdrt describes how a high-dimensional data space is cut.
Each node corresponds to a subspace of an $N$-dimensional table, modeled as a
discrete hypercube, $([0, |Dom_i|), \forall i \in [0, N))$.
Each node logically holds all records that belong to its hypercube.
The root of the tree, $([0, |Dom_i|), \forall i)$, represents the whole table.
We assume that the domain of each dimension is known, and its attribute values are in $[0, |Dom_i|)$.

% {\bf Notation.} We model an $N$-dimensional relation as a discrete hypercube, represented by $([0, |Dom_i|), \forall i \in [0, N))$.  We assume that the domain of each dimension (column) is known, and its attribute values are in $[0, |Dom_i|)$.
% , written as $([l_i, r_i), \forall i \in [0, N))$.  It logically holds all records that belong to this hypercube.  The root of the tree, $([0, |Dom_i|), \forall i)$, represents the whole table.

Each internal node $n$ has two children, where the left child satisfies a particular predicate $p$---attached to node $n$---and the right child satisfies $\neg p$.  For now, we assume each predicate to be a simple unary form, \textsf{(attr, op, literal)}, where \textsf{op} is a numeric comparison, but the framework supports arbitrary predicates as well.  We call predicate $p$ a \textit{cut} on node $n$. 

\qdrt differs from the classical \emph{k}-d tree~\cite{bentley1975multidimensional}.  \emph{k}-d tree can be seen as a simple form of \qdrt, in that they typically come with heuristics such as assuming cuts to be unary, cuts alternating among dimensions, and cuts points chosen as each dimension's median value.  \qdrt does not assume these construction heuristics.

{\bf Example.} Figure~\ref{fig:tree-example} shows an example \qdrt on two columns, \textsf{(cpu,mem)}.  The root is cut with predicate $cpu < 10\%$.  The resultant two children are cut with $mem = 10GB$ and $cpu < 5\%$, respectively. In our implementation, the literals, e.g., ``$10\%$'', are dictionary-encoded as integers.

We next describe the usage of \qdrt in data routing and query processing.  We later present algorithms to construct \qdrt in Sections~\ref{sec:greedy} and~\ref{sec:rl}.

\subsection{Routing Data}
\label{sec:data-routing}

Our overall strategy is to use a \qdrt to assign data to blocks on storage. The routing of data to blocks is carried out as follows.
% Physical partitioning is carried out by consulting a completed \qdrt. 
Each record ``arrives'' at the root and is recursively routed down.  
At each node, the tagged predicate $p$ is evaluated; if $p(record)$ is true, it is routed to the left, otherwise to the right.  Each record uniquely lands in a leaf due to the binary split ($p$ or $\neg p$). Each leaf thus represents a set of physical blocks to be persisted. \revision{Records are stored with an additional block ID ({\tt BID}) field to denote the block they belong to, and the dataset is partitioned by this field.}

In practice, we route large batches of records at a time, taking advantage of vectorized instructions.  Further, threads can load different batches of records in parallel (assuming the appends at the leaves are protected with locking).

\subsection{Semantic Description of Nodes}

\begin{table}[t]\centering \small%
\begin{tabular}{@{} l p{5cm} @{}} \toprule
\textbf{Fields of node $n$} & \textbf{{Definition}}  \\ \midrule
$\texttt{n.range}$  & Hypercube describing the node's subspace. $2N$-dimensional array. \\ 
$\texttt{n.categorical\_mask}$  & Map: categorical column $i$'s name $\rightarrow$ $|Dom_i|$-dim of bits. 0 means that value is not present.  \\ 
% $\texttt{n.preds\_mask}$  & $|P|$-dim of bits.  0 means that special predicate evalutes to false for this subspace.  \\
\bottomrule
\end{tabular}
\vspace{0.25em}
\caption{\small{Semantic description of a \qdrt node.}\label{table:node-metadata}}
\vspace{-.2in}
\end{table}

As mentioned above, each allowed cut (predicate) is of the form \textsf{(attr, op, literal)}.  We allow each operator to be range comparisons, $\{ <, \leq, >, \geq \}$, or equality comparisons, $\{=, \textsf{IN}\}$.  We now describe what node metadata we need to store to process each cut.  Table~\ref{table:node-metadata}
 presents a summary. 
 
\textbf{Handling range comparisons} is straightforward, as we only need to \textit{restrict} a parent's hypercube description.  For example, Figure~\ref{fig:tree-example}'s root node has the hypercube 
\[
\textsf{root.range:} \quad [0,MAX_{cpu}),[0,MAX_{mem}),
\]
and the cut on this node, $cpu < 10\%$, produces two restricted versions for its left and right child, 
\begin{align*}
\textsf{left.range:} \quad &[0,10\%),[0,MAX_{mem})\\
\textsf{right.range:} \quad & [10\%,MAX_{cpu}),[0,MAX_{mem})  
\end{align*}

\textbf{Handling equality comparisons}, i.e., $=$ and \textsf{IN}, requires storing additional metadata.  We assume that these predicates are only issued to categorical columns.  Each node stores, for each categorical column $i$, a $|Dom_i|$-dimensional bit vector, representing the distinct values of this column.  If 1 is present at a position, the value that corresponds to that position may appear under the node's subspace; otherwise, if 0, that value definitively does not appear. It is then straightforward to process $=$ and \textsf{IN} cuts, by simply keeping (for the left child, since it satisfies the cut) or zeroing out (for the right child) the corresponding slots in the bit vector.  For example, consider a categorical column, ${priority} \in \{LOW, MED, HIGH\}$.  The root is initialized with
\[
\textsf{root.categorical\_mask:} \quad (priority \rightarrow [1, 1, 1])
\]
since any of the three values may potentially appear.  If we cut the root with $priority = MED$ (say, second value in the ordered domain), the left and right child would have the following categorical masks:
\begin{align*}
\textsf{left.categorical\_mask:} \quad & (priority \rightarrow [1, 1, 1])\\
\textsf{right.categorical\_mask:} \quad & (priority \rightarrow [1, 0, 1])
\end{align*}
because the right child must satisfy $\neg(priority = MED)$.
Overall, this scheme is similar to ``dictionary filtering'' in popular persistent formats such as Parquet.

Thus, $(\textsf{range}, \textsf{categorical\_mask})$ make up a node's semantic description.  We make an optimization in the case of when data has fully been routed down a \qdrt.  In this scenario, we can freeze the tree and replace each leaf's \textsf{range} with a min-max index over the leaf's records.  The min-max index serves to ``tighten'' the range hypercube.

Each node of the \qdrt has a semantic description as described. Further, based on our routing strategy, the blocks assigned to each leaf together have the completeness property, i.e., for a given leaf's semantic description, every record satisfying the description is stored at the corresponding leaf.

\subsection{Query Processing}
\label{sec:query-proc}
\removedfromrevision{
\qdrts allow an arbitrary query to obtain the list of data blocks that it may
need.
In the most general case, it can always return all leaves---this is
equivalent to asking the compute engine to perform a full scan.
However, we can often do better than full scan: we intersect a query with each
\qdrt leaf to remove data blocks it definitely does not need.
% often take advantage of specific predicates in the query to
% significantly shorten the list.
% The downstream scan operator then needs only fetch a subset of blocks.
The list of required data blocks is passed to the underlying physical layer for retrieval.
}

\revision{
A simple way to process queries is to directly execute them on a dataset partitioned 
by the block ID ({\tt BID}) field introduced by \qdrt (Sec.~\ref{sec:data-routing}). 
% A simple way to process queries is to directly execute them on the dataset laid out by a \qdrt.
In this case, which requires no intervention during query processing, the traditional partition-pruning~\cite{moerkotte1998small, graefe2009fast, dageville2016snowflake} block-level indexes (e.g., min-max) are used for actual block-skipping on a best-effort basis. For further effectiveness, we instead intercept queries submitted by users and augment them to effectively use \qdrt for partition pruning as follows. Queries are routed through the \qdrt and augmented with a \textsf{BID IN (...)} clause that lists the pruned set of block IDs. Modern databases can use this explicit predicate to prune blocks, without modifications to the database internals. If desired, the query routing functionality can also be integrated into the DBMS to make the process entirely transparent.}

\revision{To obtain the {\tt BID} list}, we loop over each leaf description, check whether the query
logically intersects with the leaf subspace, and return the IDs of all
intersecting leaves.  Concretely, for any (unary) range predicate (recall from
last section, these include $\{ <, \leq, >, \geq \}$), we perform a simple
interval intersection check against each $\textsf{leaf.range}$.  For any
equality predicate ($=$ and \textsf{IN}), we check the corresponding bit vector
slot in $\textsf{leaf.categorical\_mask}$.
Alternatively, we could also ``route'' the query down the tree to reach a set
of leaves; however, we find scanning leaf metadata to be efficient enough,
especially when leaf metadata is grouped together for fast access.

Building on top of checks for \emph{predicates}, the intersection checks for
\emph{queries} are natural extensions.  We allow a query to be arbitrary conjunction or
disjunction of unary predicates (and of lower-level conjuncts/disjuncts).  The
intersection logic for \textsf{AND} is simply that it intersects if all
of its conjuncts do.  Likewise, an \textsf{OR} intersects if any of its
disjuncts does.

% This check
% is done just before the scan operator, and the list of required data blocks is
% passed to the underlying physical layer for retrieval.

% Talk about conjunct, disjunct.

% We consider the compute engine (e.g., SparkSQL) that orchestrates
% than orthogonal choice.

% This is achieved either by routing the query down the tree (checking whether the
% query hypercube intersects with each child node's hypercube), or, more simply,
% by scanning all leaf hypercubes and testing for intersection.  The latter
% approach is efficient

% Consider a leaf block of a \qdrt with some hypercube $([l_i, r_i), \forall i)$. A query can safely skip this block, if the query hypercube does not intersect with this hypercube.

% {\bf Supported queries.}  Any queries are supported.

% Talk about goodness of a tree.

% For ease of exposition, we restrict attention to (1) binary tree shapes, and (2) the two children of any internal node partitions the space of their parents. 

\subsection{Choosing Candidate Cuts}
\label{sec:allowed-cuts}
Prior to discussing algorithms to construct \qdrt, we describe choosing the set of allowed cuts.
This set serves as the search space for the construction algorithms.

We opt for a simple treatment.  Since we are given a target workload $W$ of queries, we simply parse them through a standard SQL planner and take \emph{all pushed-down unary predicates} as allowed cuts.
% For example, from a target query \texttt{SELECT ... WHERE (R.a < 10 OR R.b > 90) AND (R.c IN (0, 4))}, three allowed cuts are extracted: (1) \textsf{R.a < 10}, (2) \textsf{R.b > 90}, and (3) \textsf{R.c IN (0, 4)}.
For example, from a target query:
\begin{verbatim}
SELECT ... FROM R
WHERE (R.a < 10 OR R.b > 90) AND (R.c IN (0,4))
\end{verbatim}
three cuts are extracted: (1) \texttt{R.a < 10}, (2) \texttt{R.b > 90}, and (3) \texttt{R.c IN (0,4)}.
We find that our algorithms can easily handle a few hundreds to low thousands of candidate cuts.
% LocalWords:  Dom unary attr sep mem

 \section{Greedy Construction of \qdrt}\label{sec:greedy}
% \subsection{Algorithm}

\revision{The construction of a \qdrt is an NP-hard combinatorial optimization problem. 
% Greedy algorithms are typically used to solve such problems; 
Greedy algorithms are a typical family of solutions 
that are usually efficient and make locally optimal choices. Hence, we start by proposing} a greedy algorithm to construct the \qdrt. 
 We begin with all the tuples in a single block, i.e., the \qdrt has a single root node that contains all the tuples. In each iteration, we split a leaf node whose size is larger than $2b$ into two child nodes, and make sure the two children have size at least $b$. When choosing the cut for a node, we use the one that maximizes $C(T)$, i.e., the number of tuples skipped by the partitioning $\mathcal{P}_T$ induced by \qdrt $T$. The idea is similar to decision tree construction, except that in decision tree learning, the predicate is chosen using a different criterion such as information gain. 
 
 To present the algorithm, we define an action $a=(p,n)$ as applying cut $p$ to node $n$ in a \qdrt $T$. The result of action $a$ is denoted as $T\oplus a=T\oplus (p,n)$. In $T\oplus a$, node $n$ becomes the parent of two child nodes: the left child $n^p$  contains all the tuples in $n$ satisfying $p$, and the right child $n^{\neg p}$ contains all the tuples in $n$ satisfying $\neg p$. 
%  \todo{Before we go into formalism and proofs, should we discuss pseudocode of the greedy algorithm?  The above sentence ``When choosing the cut...'' may be too succinct of an algorithmic description?}
 \begin{algorithm}
 \caption{Greedy construction of \qdrt}
 \label{algo:greedy}
 \begin{algorithmic}
 \State \textbf{Input}: Tuple set $V$, min block size $b$, workload $W$, candidate cut set $P$
%  \Output{\qdrt $T$}
 \State \textbf{Initializatioin}: Set $T_0\gets V, t\gets 1$, $CanSplit \gets True $
 \While{$CanSplit$}
    \State $CanSplit \gets False $
    \For{each node $n\in T_{t-1}$ on the last level}
        \If{$n.size\ge 2b$} 
            \State $p\gets\arg\max_{p\in P,|n^p|\ge b, |n^{\neg p}|\ge b}C(T_{t-1}\oplus(p,n))$
            \If{$C(T_{t-1}\oplus(p,n))>C(T_{t-1})$}
                \State $T_{t}\gets T_{t-1}\oplus(p,n)$
                \State $t\gets t+1, CanSplit \gets True $
            \EndIf
        \EndIf
    \EndFor
 \EndWhile
 Return $T_{t-1}$
 \end{algorithmic}
 \end{algorithm}
 
 Our algorithm is presented in Algorithm~\ref{algo:greedy}. The main computation is to choose the cut $p$ that maximizes the greedy criterion $C(T_{t-1}\oplus(p,n))$ for each node $n$. For each level of the tree, the cost to executed the for loop is bounded by $O(|V||P|)$. The total cost of the while loop is bounded by $O(|V||P|d)$, where $d$ is the final depth of the tree. $\log_2 \frac{|V|}{b}\le d<|V|/b$. In the worst case, when the tree is least balanced, the complexity is quadratic to $|V|$. In the balanced case, the complexity grows as $|V|\log |V| $. If the number of partitions $|V|/b$ is a constant, then the cost is linear to $|V|$. 
 
 \subsection*{Approximation Guarantee}
 Under certain generic assumptions, we can prove that the greedy algorithm has a multiplicative offline approximation guarantee, and an additive online approximation guarantee. To the best of our knowledge, this the first algorithm with such guarantees. Our proof technique defines a new notion of submodularity: tree submodularity. We are unaware of notions of similar kind in other tree construction algorithms.
 
 Let $t$ be the number of leaf nodes in a \qdrt. The total number of nodes is $2t-1$. While the result of the greedy algorithm is invariant to the order of leaf splitting (which leaf to split at each iteration), we fix the order to be top-down, left-to-right for ease of analysis. In each iteration $i$, a cut $p_i$ is chosen for the leftmost node $n_i$ on level $l_i$. We define this action as $a_i=(p_i,n_i)$. 
To characterize the properties of $C(T)$, it is useful to define an encoding of the \qdrt.
% two encodings of the \qdrt.
\removedfromrevision{
 \begin{definition}[Level-wise Encoding]
 A level-wise encoding $E(T)=\{n_i\}_1^{2t-1}$ of a \qdrt indexes all the nodes in a top-down, left-to-right order: $\forall 1\le i<j\le 2t-1$, node $n_i$ is either on top of node $j$, or on the left of node $j$ at the same level.
 \end{definition}
 }
 \begin{definition}[Action Encoding]
 An action encoding $A(T)=\{a_i\}_1^{t-1}$ of a \qdrt is the sequence of cuts chosen following the top-down, left-to-right order. We denote the tree after iteration $r$ as $T_r$, and $T_r=T_0\oplus a_1\oplus \dots \oplus a_r=T_0\oplus A_r(T)$. 
 \end{definition}
  After applying action $a_i$, a leaf node is split into two child nodes.
%   , which are appended to the end of the level-wise encoding. 
  By definition, $T=T_{t-1}=T_0\oplus A_{t-1}(T)$.
 \removedfromrevision{
  \begin{lemma}[Tail Monotonicity]\label{lemma:monotonic}
  Given two \qdrt $T$ and $T'$, if $E(T)$ is a prefix of $E(T')$, then $C(T)\le C(T')$.
 \end{lemma}
 Lemma~\ref{lemma:monotonic} ensures that during the construction of \qdrt, the objective keeps increasing. 
 }
 \removedfromrevision{
 %  To further characterize the structure of $C(T)$, 
 Further,
 we define the following notion of concatenation of two trees.
 \begin{definition}[Action-based tree concatenation]
 Given two \qdrt $T$ (with $t$ leaf nodes) and $T'$, the action-based tree concatenation $T\oplus T'$ is defined as replacing the leftmost leaf node $n_t$ on level $l_{t}$ in $T$ with a subtree $n_t\oplus A(T')$. That is, the concatenation is obtained by applying the sequence of cuts in $T'$ to the next leaf node to split in $T$.
 \end{definition}
 \begin{lemma}[Head Monotonicity]\label{lemma:head}
 Given two \qdrt $T$ (with $t$ leaf nodes) and $T'$, $C(T')\le C\left(T\oplus_{i=1}^t T' \right)$.
 \end{lemma}
 This lemma is true because every leaf node in $T\oplus_{i=1}^t T'$ contains a subset of tuples of a leaf node in $T'$.
 }
 \begin{definition}[Tree Submodularity]\label{def:submodular}
 Given two nodes $n$ and $n'$ in \qdrt $T$, and actions $a=(p,n),a'=(p,n')$. If $n'$ is an ancestor of $n$, let $T'$ be $T$ minus all descendants of $n'$. We say a \qdrt space is tree-submodular if for any such $T,n,n',a$ in the space, $C(T\oplus a)-C(T)\le C(T'\oplus a')-C(T')$.
 \end{definition}
 This property means that applying a cut in \qdrt has a diminishing return as the tree grows deeper. Let $Q(p)\subseteq W$ be the set of queries that can be skipped by $p$. We have the following sufficient condition for the \qdrt space to be tree-submodular. 
 \begin{lemma}\label{lemma:submodular}
 The \qdrt space is tree-submodular if the conjunction of cuts $p_1$ and $p_2$ cannot skip any query besides $Q(p_1)\cup Q(p_2)$ for all candidate cuts $p_1$ and $p_2$.
%  for any cut $p$ and query $q\in W$ such that $p$ cannot skip $q$ (i.e., $q$ intersects with both $p$ and $\neg p$), there does not exist a set of cuts $P$ such that $\bigwedge_{p'\in P}p'$ cannot skip $q$ and $\bigwedge_{p'\in P}p'\bigwedge p$ can skip $q$.
 \end{lemma}
 For example, if the workload $W$ only consists of conjunctive range queries, and each cut is a range predicate, the above condition is satisfied. 
%  For space reasons, we defer the proof to a technical report.
%  The proof is removed from the revision and will be placed in a technical report.
 \removedfromrevision{
 \begin{proof}
 Let $n'$ be an ancestor node of $n$. We define the set of queries that can be skipped by $n$ as $Q_n\subseteq W$. It is easy to see $Q_{n'}\subseteq Q_{n}$. Let $Q(n,p)$ and $Q(n,\neg p)$ be the sets of queries skipped by the left and right child of $n$, and $c(n,p)$ be the number of tuples in node $n$ satisfying $p$. We then have:
 \begin{align}
     &C(T\oplus a)-C(T) \nonumber\\
                       =&c(n,p)|Q(n,p)\setminus Q_n|  +c(n,\neg p)|Q(n,\neg p)\setminus Q_n|\nonumber\\
                       =&c(n,p)|Q(p)\setminus Q_n|    +c(n,\neg p)|Q(\neg p)\setminus Q_n|\label{eq:sub1}\\
                    \le &c(n',p)|Q(p)\setminus Q_{n'}|+c(n',\neg p)|Q(\neg p)\setminus Q_{n'}|\\
                       =&c(n',p)|Q(n',p)\setminus Q_n|+c(n',\neg p)|Q(n',\neg p)\setminus Q_n|\label{eq:sub2}\\
                       =&C(T'\oplus a')-C(T')\nonumber
 \end{align}
 Both Eq.~\eqref{eq:sub1} and~\eqref{eq:sub2} use the sufficient condition.
 \end{proof}
 }
 
 \begin{theorem}
 If the \qdrt space is tree-submodular, the greedy top-down construction algorithm produces a \qdrt $T$ whose overall skipping capacity $C(T)$ is no worse than: 
 
 (a) $OPT-\frac{2|V|}{b}\left(C(T)-C(T^{-1})\right)$
 
 (b) $\left(1-\frac{b}{|V|}^{\frac{b\log_2 e}{2|V|}}\right) OPT$
 
 where $OPT$ refers to the skipping capacity of the optimal \qdrt $T^*$, and $T^{-1}$ refers to the sub \qdrt in $T$ by removing all the leaf nodes.
 \end{theorem}
 Bound (a) is an online bound because it depends on the algorithm output $C(T)$ and $C(T^{-1})$. Bound (b) is an offline bound independent of the algorithm output.
%   The proof is removed from revision and will be cited as a technical report.
 For space reasons, we defer the proofs to our technical report~\cite{qdtree}.

\section{\qdt using Deep RL}
\label{sec:rl}

\revision{The greedy technique presented above makes locally optimal choices which may lead to global suboptimality. At the other extreme is dynamic programming (DP), or equivalently, memoized search. It can find the optimal solution, but is infeasible given our large high-dimensional search space, leading to a need for approximate DP~\cite{powell2007approximate, powell2016perspectives}. In this paper, we propose leveraging deep RL to perform an approximate, accelerated, and incremental memoized search.

\sys is our deep RL agent that constructs routing trees optimized for a target dataset and workload.  At a high level, the algorithm repeatedly constructs many trees, initially making random cuts (i.e., randomly sampling a tree from the set of all valid trees), and gradually learns to identify better cuts through rewards.  After attempting a fixed number of trees or if a timeout is reached, the best tree found is deployed. This approach brings several key benefits:
\begin{itemize}[leftmargin=*]
\item Instead of remembering all the exact search states and the optimality (reward) for them, it featurizes the states and uses a model to predict the reward under the states.
\item Instead of enumerating all the follow-up actions of a search state to observe the reward from each of them, it samples a subset of such actions and updates the model from the observed rewards.
\item It can incrementally produce better trees, letting us deploy solutions quickly based on time or CPU budgets.
\end{itemize}

We next start with motivating arguments to illustrate the above intuition, and then present \sys in detail.}

\subsection{Motivation for RL}
\label{sec:why-rl}
% \subsection{Challenges in Routing Tree Construction}

Routing tree construction presents several unique challenges that we argue are good fit for RL.

First, the exact goodness of a tree is only measurable after the whole tree is
completed.  Typically, a tree is completed after dozens or hundreds of
cuts.  Thus, when deciding what cut to make, we either
approximate its benefit \emph{at that single step} (e.g., a greedy criterion), or we randomly sample a cut
from some (learned, gradually refined) distribution, and then accurately
attribute benefits of each decision once the true goodness is calculated.  We
will show long-term consideration leads to higher quality trees than greedy
consideration.  RL methods are thus a natural fit because they study
the optimization of long-term, cumulative rewards.

\begin{figure}[t!]
\centering
\includegraphics[width=.75\columnwidth]{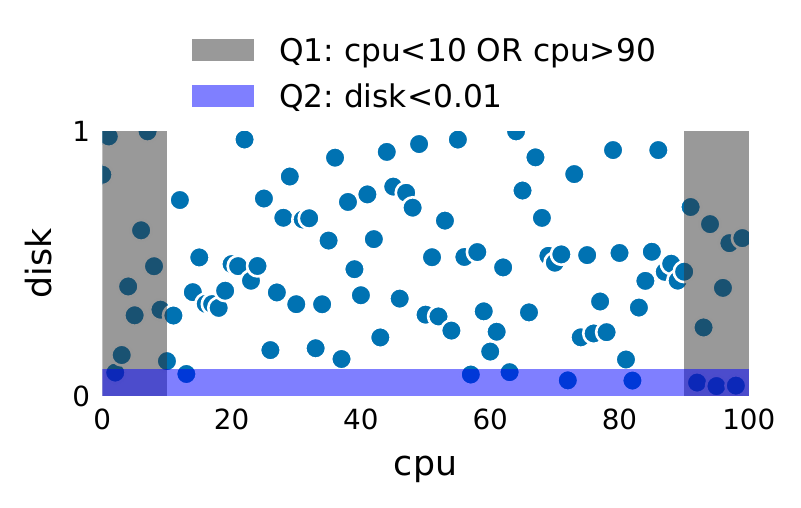}
\vspace{-.1in}
\caption{\small A dataset with disjunctive queries. Regions selected by Q1/Q2 are shown in grey/blue. The candidate cuts are: \{cpu<10, cpu>90, disk<0.01\}.  The first two cuts cannot skip any query, so Greedy opts for the third cut, resulting in a scan ratio of 50.5\%.  \sys is not limited by the forms of queries; it produces a layout with a scan ratio of 10.4\%, a 4.8$\times$ improvement. Discussion in Section~\ref{sec:why-rl}.\label{fig:disjunctive}}
%\vspace{-.4in}
\end{figure}

Second, an RL method does not make assumptions on the query or data distribution.
It requires only a black-box learning signal, the skipping quality of a tree.
To be concrete, we now present a microbenchmark to showcase the potential advantage due to this generality.

In Figure~\ref{fig:disjunctive}, we plot a simple dataset with two columns, \textsf{(cpu, disk)}.
We draw $cpu \sim \text{Unif}[0, 100)$ and $disk \sim \text{Unif}[0, 1)$.
Query 1 is a disjunctive query on $cpu$ (perhaps looking for anomalies at either ends), and Query 2 is a unary filter on $disk$.
Recall from last section that the optimality proofs of our Greedy construction relies on the tree-submodularity. When the query workload contains disjunctive range queries and the candidate cuts are only simple range predicates, tree submodularity is not satisfied.
Our Greedy algorithm is forced to choose the cut on $disk$---since, the two cuts on $cpu$ provide zero skipping capability (making either cut cannot skip Q1 nor Q2), whereas the cut on $disk$ provides a non-zero gain.
% within a single step, it provides a better (non-zero) skipping gain than the other two cuts.
This results in a layout of the following two blocks:
\begin{itemize}
  \item Block 1: $disk < 0.01$
  \item Block 2: $disk \geq 0.01$
\end{itemize}
Thus, Q1 has to scan the large portion of unselected records in the middle.
Our deep RL agent, \sys, is able to produce a $4.8\times$ better partitioning:
\begin{itemize}
  \item Block 1: $disk < 0.01$
  \item Block 2: $(disk \geq 0.01) \land (cpu > 90)$
  \item Block 3: $(disk \geq 0.01) \land (cpu < 10)$
  \item Block 4: $(disk \geq 0.01) \land (cpu \leq 90) \land (cpu \geq 10)$
\end{itemize}
Hence, under this layout, both Q1 and Q2 can skip block 4, which contains a majority of records.
This showcases the power of RL as a black-box optimization method.

Lastly, a large search space needs to be navigated.  The number of candidate
cuts can be large, potentially $O(100)$ or $O(1000)$.  Further, the number of
data dimensions can be in the dozens or hundreds.  Deep RL (compared to
classical RL) methods have shown successes in tackling such high-dimensional problems
For instance, OpenAI Five~\cite{OpenAI_dota}, a deep RL agent for successfully playing Dota, considers an action space of $O(1000)$ dimensions and an $O(20000)$-dimensional observation space.
Our \sys agent uses the same scalable learning algorithm as OpenAI Five, taking advantage of recent algorithmic advances.
% For instance, AlphaStar, a deep RL agent for successfully playing StarCraft II, considers a space of $\sim 10^{26}$ actions.
% make use of neural networks as high-capacity function
% approximators, which can learn to compactly encode correlations between
% different cuts and data hypercubes.

% When combined, these challenges call for a deep reinforcement learning approach.
% Still, there is a limit in current state-of-the-art deep RL applications: for
% instance, AlphaStar~\cite{alphastar} considers ?? actions only, whereas our agent
% has comparable or bigger action space.  %We next describe algorithmic details.

% \todo{Highlight the fact it doesn't make assumption on the form of queries---a
%   more general algorithm.  Only a black-box skipping ratio calculation is required.  Show example.}
% Data: column A ~ [0, 100] integers uniformly distributed, column B ~ Unif[0, 1) float. Target workload: Q1 = (a < 10 OR a > 90), Q2 = ( b < 0.01 ). TD scan ratio: 50.5. RL scan ratio: 10.6.   -- roughly 5x less.

We next describe the detailed design of \sys, a deep RL agent that learns to construct \qdrts.

% \subsection{\sys: \qdrt via deep RL}
\subsection{\sys: the Deep RL agent}

To apply any RL algorithm, we first need to define the tree construction Markov Decision Process (MDP).
The \textbf{state space}, $S$, is defined to be any subspace of the entire data space of the relation under optimization.
The \textbf{action space}, $A$, is the set of allowed cuts.
Taking an action (cut) on a state (node) produces two new states, which we append into a queue for exploration.
The queue is initialized with a root state (the root node) when starting each tree construction.

% We define the state space $S$ to be any subspace of the entire data space.  We
% define the action space $A$ to be a discrete categorical distribution over all
% allowed cuts.
% We now explain in detail.

The \sys agent, at its core, consists of two learnable networks parameterized by $\theta$:
(1) the \emph{policy network}, $\pi_\theta: S \rightarrow A$, takes a state and emits a probability distribution over the action space (``given a node, how good are the cuts?''), and
(2) the \emph{value network}, $V_\theta: S \rightarrow \mathbb{R}$, estimates the expected cumulative reward from a given state.

Sequentially, the agent (1) takes a node $n$ off the exploration queue,
(2) evaluates its current policy, $\pi_\theta(n)$,
(3) samples an action from this output distribution,
(4) applies the sampled action (cut) on node $n$ to produce new nodes.
We use Proximal Policy Optimization (PPO)~\cite{ppo} as the underlying learning algorithm, a variation in the policy gradient family of methods.
This update rule is used as a black-box subroutine and is not fundamental to the design of \sys.

{\bf Intuition.} %An  refers to the construction of one tree.
We start each \emph{episode} (the construction of one tree) with the root state (the singleton tree with a root node).
The agent takes actions and transition into next state(s).
Once a stopping condition is reached, described next, the episode is ended and we obtain a completed \qdrt.
We calculate the reward of this episode, i.e., the data skipping ratio achieved by the tree, and invoke PPO for gradient updates to $\theta$.
With the updated behaviors, the agent starts the next episode.
Through repeatedly constructing \qdrt, the RL agent becomes better.
At the beginning of training, a randomly initialized $\theta$ implies random behavior: random cuts are made and the skipping ratios would not excel.
However, as more trees (episodes) are explored, the refined $\theta$ is encouraged to make cuts that achieve higher skipping ratios.

We next discuss algorithmic details.

\subsubsection{Stopping Condition}
To prevent an endless sequence of cuts, we must define an appropriate stopping condition.
A naive condition would be stopping after a pre-determined number of cuts are made.
This is problematic, because we do not know a priori the number of cuts required to achieve good data skipping for a given dataset-workload.

Instead, we connect back to Problem~\ref{def:maxskip}'s requirement where each leaf block must contain at least $b$ records.
The agent is allowed to make a cut $p$ on current node $n$, if the resultant children (approximately) contain more than $b$ records.

This approximation is achieved by testing the cut on a data sample.
First, at algorithm initialization, we take a data sample of ratio $s$ from the dataset (we find $s=0.1\%$ to $s=1\%$ generally work well).
All episodes (the exploration of trying out different trees) reuse the fixed sample.
The data sample is assigned to the root node.
Next, as a cut is made, we evaluate the cut on the data sample, and obtain a subset of records that satisfy it and a subset that does not.
If both have more than $s \cdot b$ records, we call the cut legal and allow it to proceed.
Lastly, if a current node $n$ has no legal cuts in action space, we stop cutting on it and form a leaf.

\subsubsection{Reward Calculation}
After a tree $T$ is produced, \sys calculates rewards for all actions taken.  The rewards serve as important learning signals: they allow the agent to learn to distinguish profitable cuts.

% We define the reward function $R(n)$ for every intermediate node $n$ as follows:
First, we define $S(n)$, the number of skipped records under node $n$ across all queries:
\[
 S(n) :=
  \begin{cases}
    C(\textsf{n.records}) & \text{if } n \text{ is a leaf} \\
   % \text{records of $n$ skipped by all queries (Sec.~\ref{sec:query-proc})}
   S(\textsf{n.left}) + S(\textsf{n.left})       & \text{otherwise}
  \end{cases}
\]
Recall from Equation~\ref{eq:c-func} that $C(\cdot)$ refers to the number of records skipped across the workload.  Here, $\textsf{n.records}$ is the set of records routed to node $n$ during tree construction---thus a subset of the small data sample we take.  Since the sample is small, this ensures reward calculation is efficient.

We then assign a reward $R$ for every action taken, i.e., for every intermediate node $n$ and corresponding cut $p$:
\[
 R((n, p))  := S(n) / (|W| \cdot |\textsf{n.records}|)
\]
Namely, we normalize number of skipped records under $n$ to $[0,1]$ by scaling with $1/(|W| \cdot |\textsf{n.records}|)$, where the denominator is the maximum number of skipped records possible under $n$ (the best case of all queries skipping all its records).
The PPO update rule is then invoked with a list of state-action-reward tuples.

\subsubsection{Implementation}
This section discusses detailed implementation of the networks.
The policy network $\pi_\theta$ and the value network $V_\theta$ have shared weights,
which are two fully-connected layers, 512 units each, with ReLU activation.
Each network has its own output layer: for the policy network, it is a $|A|$-dimensional linear projection; for the value network it is a scalar projection.

Each state is featurized as the concatenation of $\texttt{n.range}$ and $\texttt{n.categorical\_mask}$. Due to the potentially large values in the former component, we binary-encode both vectors (i.e., these vectors are encoded in bits).  The action space is a discrete categorical distribution, usually with a dimensionality in a few hundreds.

We found that neural network computation is not a bottleneck in our setup.  Routing records and calculation of rewards, on the other hand, take up a significant portion of tree construction time.  We therefore only use CPUs for our agent (although using a GPU for neural network computation yields a slight overall speedup).

% \todo{approximate layout constraint? System diagram? Algorithm pseudocode box?}

\subsubsection{Related Work}
\label{sec:rl-related}
Our formulation of using deep RL to learn a tree under custom quality metrics is inspired by
NeuroCuts~\cite{neurocuts}, a deep RL algorithm to construct packet classification trees. We compare the two work below.
% in the networking community.  We discuss similarities and differences below.

First, \sys adopts their overall approach of tree-structured MDP: each node $n$ is treated as an independent state, and receives a normalized reward as if the agent is asked to start constructing a tree from that node.  This means the agent is tasked to solve each subproblem independently.

Second, while NeuroCuts assumes no knowledge of workload, \sys optimizes a target data-workload pair.  This affects the choice of the actions.  Their action space includes generic actions (e.g., ``cut dim K in N equal parts'') while we obtain our actions from the unary predicates of the target workload.  We find that such generic actions do not make sense in our setting---the domain of an attribute can be large, and cutting at non query-aligned literals is suboptimal.

Third, we make specific optimizations for our data analytics setting.  A data sample is required for us to calculate invalid cuts and respect the layout constraints.  We also propose special treatment of categorical predicates in featurization.

\revision{
We find our RL-based solution to already achieve within $20$\% of the true dataset selectivity, which is itself a lower bound for the optimal solution, for the TPC-H benchmark (Sec.~\ref{sec:eval}). Therefore, we do not consider alternate DP approximation or optimization techniques~\cite{powell2007approximate, powell2016perspectives} in this work.
}

% LocalWords:  NeuroCuts

\section{Framework Extensions}
\label{sec:opt}

Having described two algorithms to construct a \qdrt, we are now in a position to
discuss extensions to our framework.

\subsection{Advanced Cuts}
\label{sec:adv-cuts}
Thus far we have assumed the candidate cuts are single-column predicates (Section~\ref{sec:allowed-cuts}).
They are desirable because their simplicity allows for fast evaluation during tree construction.
Nevertheless, \qdrt can be extended to support binary cuts of the form
\textsf{(attr1, op, attr2)}. Recall from Table~\ref{table:node-metadata} there are two
components of a node's semantic description, and neither of them can describe a binary cut. 
We append a new component to each node $n$'s description:%
\[
\textsf{n.adv\_cuts:} \quad \text{a bit vector of size |AC|}
\]%
where the constant $|AC|$  denotes the number of \emph{advanced cuts} to support
and is specified for each workload a priori.  Each position $i$ corresponds to
``does this node contain records that satisfy advanced cut $i$'', with zero
indicating no and one indicating potentially yes.  This is the same semantics as \textsf{categorical\_mask}.

For instance, the TPC-H workload contains non-join binary filters such as:
\begin{itemize}
  \item $AC_0$: \texttt{c\_nationkey = s\_nationkey}
  \item $AC_1$: \texttt{l\_shipdate < l\_commitdate}
  \item $AC_2$: \texttt{l\_commitdate < l\_receiptdate}
\end{itemize}
A vector of $(0, 1, 1)$ thus indicates the first condition is definitely not met
(i.e., it describes a subspace of records whose \texttt{c\_nationkey} does
\emph{not} equate \texttt{s\_nationkey}).

\removedfromrevision{
\begin{table}[htp]\centering \small%
\begin{tabular}{@{} l l l @{}} \toprule
  &  w/o advanced cuts &  all cuts  \\ \midrule
Greedy \qdrt & 4.07\% & 2.57\%    \\
RL \qdrt &  3.97\% & 2.35\% \\ \bottomrule
\end{tabular}
\caption{\small Ablation study: how search space affects the quality of layout in terms of percentages of tuples accessed (lower is better). Detailed TPC-H workload setup is discussed in Section~\ref{sec:eval-setup}. \todo{Update} \label{table:allowed-cuts}}
\end{table}

Table~\ref{table:allowed-cuts} shows an ablation study on the impact of
including these three special cuts into the search space.  We observe an up to
$1.68\times$ better quality for the layouts produced by Greedy and RL.  This makes sense,
as the three advanced cuts enlarge the search space and they are low-selectivity cuts
to start with (especially $AC_0$).
}

Lastly, the same mechanism also handles LIKE predicates or even stateless UDFs
(with the caveat that, clearly, the cost of evaluating the predicates depends on their inherent complexities).
The user can impose a limit on the maximum number of advanced cuts to support.

\subsection{Data Overlap}

With the abundance of cheap storage in the cloud, one desirable feature for an
analytics system is to trade space for potentially faster execution time.  A
fruitful line of work has dedicated to this problem, e.g., materialized views,
which we review in related work.  We now discuss how \qdrt can also naturally
support duplicating data.

\begin{figure}[t!]
\centering
\includegraphics[width=.55\columnwidth]{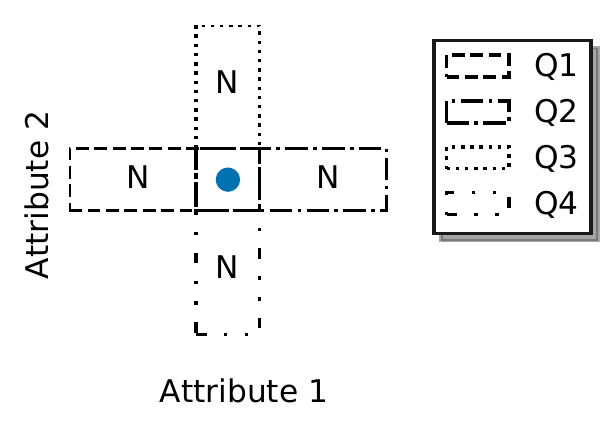}
\vspace{-.10in}
\caption{\small A scenario where significant data skipping is gained by
  replicating a single record.  Each query selects $N+1$ records. The queries
  only overlap in the one tuple placed at the center. If the space is naively
  cut in a binary fashion, 3 out of 4 queries each reads $N$ extra tuples.  By handling overlap, \qdrt
  replicates the singleton record to all four N-record regions, so no queries
  touch unnecessary records.\label{fig:propeller}}
\vspace{-.15in}
\end{figure}

Figure~\ref{fig:propeller} shows a 2D synthetic dataset and four queries.
Naively invoking either Greedy or \sys to construct a \qdrt for this
dataset-workload is suboptimal.
Any sequence of cuts---recall, the cut points are query literals, i.e., any of the edges in
the figure---would lead to 4 blocks: one with N+1 record, and three with N
records.  (This is due to the binary nature of the cuts.)  The three blocks would have to fetch the singleton record they need
from the first block.  Hence, a total of $3N$ extra tuples are read.

We extend \qdrt construction to handle such data overlap cases as follows.
Observe that the reason the ``lucky'' $(N+1)$-record block is not further cut is
due to the minimum block size constraint, $b$. We can instead launch either 
of our construction algorithms with a relaxed cutting condition: allowing one
of the children to have size smaller than $b$.  With this change the lucky block 
would be further cut into an $N$-record one and a block with the singleton record.

Once such a \qdrt is constructed, we loop through all produced leaves, and
partition them into two sets: those with size $\geq b$ and those with size $< b$
(the original constraint).  We then replicate each block in the small-size set to
any of its \emph{neighbor} blocks in the large-size set. We define
two blocks to be neighbors if their hypercubes have $N-1$ dimension boundaries
in common and the intervals at the remaining dimension are adjacent. This ensures that, with minor modifications
to node metadata, the semantic descriptions preserve completeness.
With this scheme, our algorithms could reach the optimal partitioning for the
scenario in Figure~\ref{fig:propeller} at virtually no extra storage cost.%
\vspace{-.1in}
\revision{\subsubsection{Data and Query Routing}

Data routing occurs as before, with a row routed to all matching blocks. Rows landing in a replicated block are simply copied to every replica. For query processing, the candidate
set of blocks to be considered includes all blocks that overlap with the query rectangle. However, we can leverage the completeness property of blocks to prune out blocks that
are redundant. For example, a query that asks only for the centre rectangle (with the singleton record) in the example above does not have to fetch all blocks, \emph{even though the min-max index for all four blocks includes it}. This is because of our semantic descriptions and completeness properties: we can prune away the other blocks because the first block completely covers the query rectangle. With overlap, the set of blocks scanned when evaluating a query may contain duplicate rows. To eliminate them, when scanning block ID $i$, we can simply ignore tuples that match the semantic description of any selected block with ID $< i$. A detailed evaluation of these strategies is left as future work.
}

\subsection{Data Replication: two-tree approach}

Full-copy data replication is a complimentary approach to overlap in utilizing extra storage. The extension for \qdrt construction to support such replication is natural.

First, we learn (using Greedy or RL) a \qdrt, $T_1$, optimized for the full workload $W$. Then, we can in fact build a \emph{second} \qdrt, $T_2$, tailored for the queries that experience the worst skippability under tree $T_1$. The second tree is a logical copy of the entire dataset. When constructing the second tree, we modify the reward function for RL (and the greedy criterion for greedy), by accounting for the existence of the first tree. For each query $q\in W$, we choose one of the two trees which maximizes the skippability for $q$, and calculate the number of skipped tuples $C_q$ using that tree. Then we sum up the $C_q$'s for all $q\in W$. This change naturally guides the construction of the second tree to focus on the queries with low skippability by $T_1$.
% By routing queries in $W$ through tree $T_1$ (Section~\ref{sec:query-proc}), we can collect the worst $K$ queries, say $W' \subseteq W$.  We then start producing $T_2$ for $W'$. 
Additionally, the first tree $T_1$ can be \emph{re-built} and
re-optimized with $T_2$ fixed, and iterate. Since the revised reward function keeps increasing and is upper bounded, this process eventually converges.
% As a variant, we may designate one tree as the primary data copy and prune the other tree to eliminate blocks that are not covered by any query for that tree, to save space. 
The idea may also be extended to more than two trees if needed. Exploring such extensions is a rich area for future work.

\section{Evaluation}
\label{sec:eval}

We now experimentally evaluate our designs.  We highlight key findings from the evaluation:
\begin{itemize}[leftmargin=*]
\item (Table~\ref{table:tpch-with-binary}) \qdrt-based layouts enable excellent data skipping ($1.8\times$--$61\times$
  % $3.7\times$--$51\times$
    over the state-of-the-art).
  \item Greedy construction of \qdrt produces high skipping ratios; yet, \sys, using a general-purpose RL algorithm, can still produce up to $8.5\times$ gains due to its optimization of the long-term objective.
    % agent produces up to $2\times$ additional gains.
  \item  (Sections~\ref{sec:tpc-h}, \ref{sec:errorlog}) \qdrt  provides $1.6\times$--$14\times$ physical execution speedup over a state-of-the-art baseline across execution engines, storage formats, and workloads.%
    % even on the highly-optimized columnar format, Parquet.
\end{itemize}
% \todo{These need to be updated for revision.}

\subsection{Setup and Metrics}
\label{sec:eval-setup}

We implement \qdrt as a lightweight Python library. It includes a lightweight AST and associated query-processing logic. We use the vectorized arithmetics in numpy and pandas.
\sys, the RL agent, is implemented using Ray RLlib~\cite{rllib,ray}, a scalable reinforcement learning library. \revision{We evaluate all systems using both logical and physical metrics.

{\bf Logical: Access Percentage.} We report \%tuples accessed for the whole workload achieved by various partitioners.  This metric is lower bounded by the true workload selectivity, and lower values indicate greater potential I/O savings.

{\bf Physical: Query Runtime.} We report end-to-end execution time as queries are issued over several analytics systems:}
\begin{itemize}[leftmargin=*]
\item Single-Node Spark (v2.4): All layouts execute with Spark over Parquet files stored on disk (HDD).  After a \qdrt is constructed, each leaf block is converted into a Parquet file. Other baseline layouts are stored as Parquet files as well in comparable number of blocks.
  % \revision{We execute queries both (1) with \qdrt routing, and (2) without \qdrt routing (solely depending on Parquet's metadata).}
  % As aforementioned, this takes advantage of \qdrt's optimized node descriptions (leaf hypercube and categorical masks), which Parquet translate into their min-max statistics and dictionary filters for the row groups.  SparkSQL then reads from these Parquet files.
%
% Once a \qdrt is constructed, it is converted into Parquet format where each leaf block is stored as a row group.
  % The datasets are read as appropriate Parquet files in all layouts that we evaluate (format version 1.0). %
\item \revision{Commercial DBMS: We use a commercial standalone black-box optimized DBMS. We model a scenario where partitioning (e.g., by tenant ID, query ID, or TPC-H month) is used for parallelism at a higher level, by limiting each individual query to a single degree of parallelism. We first load the datasets under different layouts into the system, which uses its own binary columnar storage format on a local SSD.}
\item \revision{Distributed Spark: We use a 4-node Spark cluster hosted on Microsoft Azure, each with 8 vCPUs, 56GB RAM, and an SSD. All data is stored as Parquet files (as with single-node Spark) on Azure Storage, a remote blob store. The \qdrt is cached at the driver for query rewrites.
}
\end{itemize}

We ensure that all layouts have a comparable number of blocks. To eliminate caching effects we clear the OS buffer cache before each query run. \revision{To evaluate the benefit of \qdrt query routing, we execute queries using explicit {\tt BID} filters (Section~\ref{sec:query-proc}; the default), or without (called \emph{no route}).} 

\subsection{Workloads}
\label{sec:eval-workloads}

% \todo{Update (1) Enlarged TPC-H; (2) other workloads here?}

We evaluate on (1) TPC-H and (2) \revision{two} real-world workloads%
% of data and queries 
from a large commercial software vendor.

% As observed by prior work~\cite{bottom_up}, in data analytics workloads it is conventional to have data accumulate in time-partitioned units (e.g., by-day or by-month), thus we apply partitioning independently and in parallel on such partitioned units.
  % what we are trying to do in the experiments is to find a table with a rich set of filters, rather than to show that we can run all tpch queries. To achieve this goal, we denormalize tpch tables to make a table that most of TPC-H queries have filters on it. In that sense, I think it’s fine if we don’t run all queries in the benchmark. (but we do run all queries that touch lineitem table).

{\bf TPC-H.} \revision{We generated TPC-H with a scale factor (SF) of 1000.} %
% \todo{clarify one month use} %% we already say "one-month" partition below
Following prior work~\cite{bottom_up}, 
% we use a scale factor (SF) of 100 and denormalize the schema.
we denormalize the TPC-H schema for the purpose of obtaining a table that many filters touch\footnote{\revision{Our technique can layout each table in the database independently using predicates over that table. Jointly optimizing the layout of multiple tables for complex join queries is left for future work.}}. 
Due to the uniform nature of TPC-H data, we apply all partitioning techniques to an one-month partition of the dataset; the month-partition totals \revision{77M tuples, 68 columns, and 85GB}.
For queries, \revision{we include \emph{all} templates that touch the \textsf{line\_item} fact table\footnote{Our goal is to find a table with a rich set of filters. To achieve this, we denormalize all tables and include all templates that touch the fact table.}. This includes} 
% we also follow their work in using 
the same 8 templates
($q_3$, $q_5$, $q_6$, $q_8$, $q_{10}$, $q_{12}$, $q_{14}$, $q_{19}$) that~\cite{bottom_up} uses, as well as \revision{7 additional templates: $q_1$, $q_4$, $q_7$, $q_9$, $q_{17}$, $q_{18}$, $q_{21}$.}
We use 10 random seeds to generate each template, resulting in a total of \revision{150}
% 80 
queries.
% The true workload selectivity is 1.2\%.
The overall scan selectivity is \revision{21.3\%}.

% \revision{
% {\bf TPC-H, SF1000.}  We also apply all techniques to an one-month partition of the denormalized SF1000 dataset (77M rows, 68 columns).  Compared to SF100, we include \emph{all} templates that touch the \textsf{line\_item} table (7 additional templates: $q_1$, $q_4$, $q_7$, $q_9$, $q_{17}$, $q_{18}$, $q_{21}$); 10 random seeds are used to generate each of the 15 templates\footnote{Our goal is to find a table with a rich set of filters. To achieve this, we denormalize all tables and include all templates that touch the fact table.}. The overall scan selectivity is $21.3\%$.  We study logical metrics on both variants and actual query performance on SF1000.
% }

{\bf Real Datasets.} \revision{Our first real dataset, called {\bf ErrorLog-Int}, consists of error logs collected from {\em internal customers} of a large software vendor.} The error logs correspond to kernel crash dump reports and are collected in real-time and loaded into a data warehouse for analysis. \revision{These error logs contain information such as a categorical event type (e.g., device crash, live kernel event, etc.) with $8$ distinct values, OS build date, OS version (string), client ingest date, and entry validity (a boolean). The dataset has $50$ columns. We collect a sample of around one week of logs, amounting to $100$ million records and $85$GB of raw data}. The dataset is also associated with a query workload imposed by automated systems via an API and users through a user interface and translated into stored procedures over the data. We extract the predicates that are pushed to storage and extract a set of \revision{$1000$} queries over $5$ dimensions that represent a majority of the workload.
The overall workload selectivity is \revision{$0.0005$\%};
% The overall true selectivity (lower bound) of the workload is \revision{$0.0005$\%};
thus, individual queries return very few results on average, usually less than $100$. All queries are of the form of {\tt IN} predicates over the categorical data, along with date ranges and {\tt LIKE} and equality predicates over the string fields.

\revision{Our second real dataset, called {\bf ErrorLog-Ext}, is also a crash dump log, but is collected from {\em external customers} (applications) around the world. This dataset is fundamentally different from the earlier one,
  collected over $15$ days with $81$ million rows ($85$GB), has more dimensions ($58$) and a much larger categorical domain of around $3600$ distinct values.
  % with a different data volume of $85$GB over $15$ days with $81$ million rows, more dimensions ($58$), and a much larger categorical domain of around $3600$ distinct values.
  We also use 1000 queries, which return more results on average, with an overall scan selectivity of $0.0697\%$.}
%   \todo{Clarify that we also have 1K queries here?}}

\begin{figure*}[t]
     \centering
     \begin{subfigure}[b]{0.49\textwidth}
         \centering
          \includegraphics[width=\columnwidth]{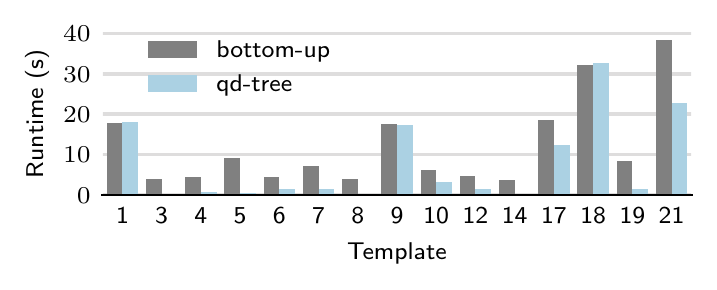}
\vspace{-.33in}
         \caption{\small{Distributed Spark}\label{fig:runtime-tpch-spark}}
     \end{subfigure}
     \begin{subfigure}[b]{0.49\textwidth}
         \centering
          \includegraphics[width=\columnwidth]{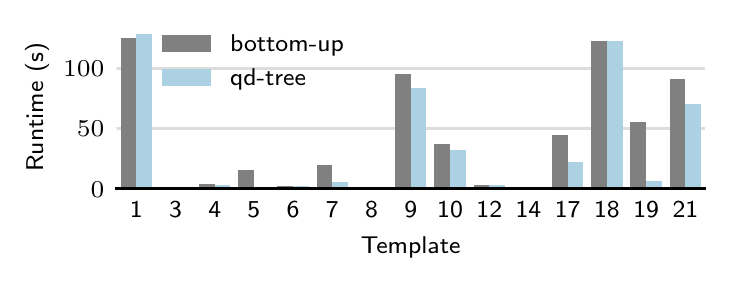}
\vspace{-.33in}
         \caption{\small{DBMS}\label{fig:runtime-tpch-dbms}}
     \end{subfigure}
\caption{\small \revision{TPC-H execution runtimes, grouped by each template.}
% \caption{\small \revision{TPC-H runtime per template. Parquet is used in (a); DBMS' own binary format is used in (b).}
\vspace{-.1in}
\label{fig:runtime-tpch}}
\end{figure*}

\subsection{Approaches}

We compare layouts produced by different algorithms, spanning from heuristics used in industry practice to the state-of-the-art approach in literature:
\begin{itemize}[leftmargin=*]
% \item Random: a random partitioner that simply shuffles the records into fixed-size blocks (except possibly the last block). \todo{Change this to Baseline and make it Random for TPC-H, Ingest for ErrorLogs?}
\item Random baseline (TPC-H): a partitioner that simply shuffles records into fixed-size blocks. %(except possibly the last block).
\item Range baseline (real workloads): range-partitioning on an ingest time column (the default scheme deployed for the real workloads).
\item Bottom-Up~\cite{bottom_up}: state-of-the-art row-grouping approach based on clustering, described in Sec.~\ref{sec:prelims}.
\item Greedy-constructed \qdrt (Sec.~\ref{sec:greedy}).
\item \sys-constructed \qdrt (Sec.~\ref{sec:rl}).
\end{itemize}
We ensure Bottom-Up, Greedy, and \sys have the same search space: the same set of candidate cuts is fed to the latter two approach, which is also fed to Bottom-Up as input to their feature selection procedure. The feature selection procedure first performs a topological sort of the features according to the subsumption relationship, and then select features one by one. The frequency of each feature is initialized as the number of queries subsumed by that feature. At each iteration, a feature not subsumed by any others is chosen, and the frequency of all the other features is discounted if the they subsume common queries with the chosen feature. A feature will not be chosen if its frequency is below a threshold.
We configure Bottom-Up to use up to 15 features, which follows the number reported in~\cite{bottom_up}. We set $b$, the minimum number of records per block, to $100$K for TPC-H and $50$K for the two ErrorLog workloads.
% For the TPC-H workload, 15 features are selected which follows the number reported in~\cite{bottom_up}.
\removedfromrevision{and 1600 unique feature vectors are produced. This matches with the numbers reported in~\cite{bottom_up}.} % # unique feature vecs are not 1600, but we cap features at 15.
\removedfromrevision{For the ErrorLog dataset, we also evaluate the commonly used partitioning scheme based on client ingest time.}

% \item Full Scan: as baseline, scanning all data for each query requires accessing 100\% of the records.

% \todo{talk about storing parquet on ssd vs. hdd vs. s3 somewhere.}

\removedfromrevision{\subsection{Results on TPC-H workload}}
% \subsection{Results on Benchmark Datasets}
% \subsection{Results on TPC-H}
\subsection{TPC-H}
\label{sec:tpc-h}

\begin{figure}[t]
     \centering
     \begin{subfigure}[b]{0.23\textwidth}
         \centering
          \includegraphics[width=\columnwidth]{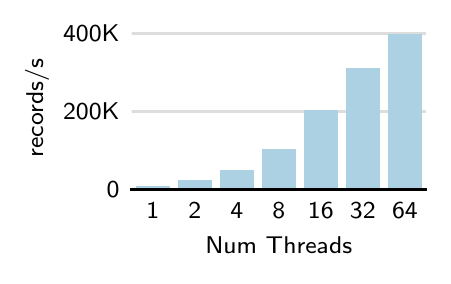}
\vspace{-.33in}
         \caption{\small{Data routing}\label{fig:data-routing}}
     \end{subfigure}
     \begin{subfigure}[b]{0.23\textwidth}
         \centering
          \includegraphics[width=\columnwidth]{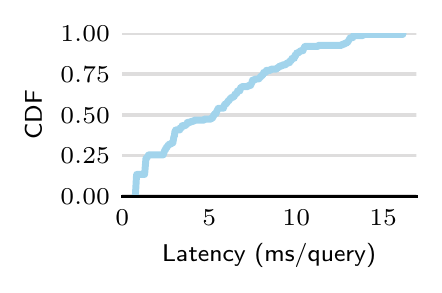}
\vspace{-.33in}
         \caption{\small{Query routing}\label{fig:query-routing}}
     \end{subfigure}
\caption{\small \revision{Performance of routing data and queries.}
\vspace{-.1in}
\label{fig:runtime-routing}}
\end{figure}

% \begin{table}[htp]\centering \small%
% \begin{tabular}{@{} l l l l @{}} \toprule
%   & \multicolumn{3}{c}{Allowed Cuts} \\
%   % \textbf{Layout}
%   &  unary cuts; min-max routing & +smart routing & + all cuts  \\ \midrule
% Top-Down & ? & ?  & ?    \\
% \sys &  ? & ? & ? \\

% \bottomrule
% \end{tabular}
% \caption{Partitioners' performance when given different allowed cuts and allowed checks.}
% \end{table}

% \begin{table}[htp]\centering \small%
% \begin{tabular}{@{} l l l l @{}} \toprule
%   &&  Top-Down & \sys   \\ \midrule
%   \multirow{2}{*}{unary cuts} & ? & ? \\
%   && & & 
% \bottomrule
% \end{tabular}
% \caption{Partitioners' performance when given different allowed cuts and allowed checks.}
% \end{table}

% Table~\ref{table:tpch} and

\begin{table}[t]\centering \small%
\begin{tabular}{@{} l l l l l @{}} \toprule
 Workload  &  Baseline  & Bottom-Up & Greedy (ours) & RL (ours)   \\ \midrule
  \removedfromrevision{SF100 & 28.8\% & 8.77\% & 2.57\% & 2.35\% \\}
% p -u partitioning_task.py --dataset=tpch-sf1000 --minsize 100000 --queries_per_template=0 --space data_pred_cut_non_atom --query_dir queries/tpch-15templates-10each-queries-distinct-1000_1009 2>&1 | tee bu-sf1000.log
% 1152 distinct vecs, 15 feats
% c_mktsegment = 0; cr_name = 0; l_discount <= 4; l_discount <= 5; l_discount <= 7; l_discount >= 8; l_receiptdate >= 91; l_returnflag = 1; l_shipdate < 91; l_shipinstruct = 1; p_brand = 15; p_brand = 16; p_brand = 17; p_brand = 18; p_brand = 19%
% bu-rowid-map-77318785records-minSize100000-1581901013.5381186.pkl
% <bottomup.BottomUpPartitioner object at 0x7f414165b048> takes 4262.5875363349915 secs
% Evaluating <class 'bottomup.BottomUpPartitioner'>
% On 77318785 records, 150 total queries: skip ratio 0.5393, access ratio 0.4607
% p -u partitioning_task.py --dataset=tpch-sf1000 --minsize 100000 --queries_per_template=0 --space data_pred_cut_non_atom --query_dir queries/tpch-15templates-10each-queries-distinct-1000_1009
%
% Evaluating <class 'partitioners.RandomPartitioner'>
% On 77318785 records, 150 total queries: skip ratio 0.4400, access ratio 0.5600
% elapsed seconds  26.805272817611694
%
% RL: we use 0.742 tree.
%
% Top Down full data tree: 1581817084.7153902.pkl
% On 77318785 records, 150 total queries: skip ratio 0.7370, access ratio 0.2630
% Top Down 1% data tree: 1581829738.648603.pkl
% On 77318785 records, 150 total queries: skip ratio 0.5598, access ratio 0.4402
TPC-H & \revision{56\%} & \revision{46.1\%} & \revision{26.3\%} & \revision{25.8\%} \\
% results/watson-nodate/other-part-chifix.log
%NOTE: TD is inconsistent in other-part.log. Using better figure.
  \revision{ErrLog-Int} & \revision{100\%} & \revision{$5.6\%^\ast$} & \revision{3.1\%} &
  \revision{0.4\%} \\
%   \revision{ErrLog-Int} & 100\% & 81.3\% & 2.09\% & 1.59\% \\

% results/watsonum-nodate/other-part-chifix.log
%NOTE: TD is inconsistent in other-part.log. Using better figure.
  \revision{ErrLog-Ext} & \revision{100\%} &  \revision{$12.2\%^\ast$} &
  \revision{$1.7\%$} &
  \revision{0.2\%} \\

%   \revision{ErrLog-Ext} & \color{red}{?}\% & 81.3\% & 2.09\% & 1.59\% \\
\bottomrule
\end{tabular}
\vspace{.05in}
\caption{\small{Logical I/O costs: percentage of tuples accessed under different layout
    schemes, compared to full scan. \emph{Baseline} is a random shuffler for TPC-H, \revision{and range partitioning on an ``ingest time'' column for the two ErrorLog workloads. ({$^*$Results of $\text{BU}^+$, our tuned version. The untuned version fares at $100\%$ and $96.9\%$, respectively}.})
    }\label{table:tpch-with-binary}}
\vspace{-.1in}
\end{table}

Table~\ref{table:tpch-with-binary} shows the percentage of tuples accessed for different layouts on the TPC-H workload. % (we set $b=100,000$, the minimum number of records per block).
Overall, \qdrt layouts provide up to 
1.8$\times$%
\removedfromrevision{3.7$\times$}
savings compared to Bottom-Up.
\revision{
\subsubsection{Physical execution}
% We test the execution runtime of TPC-H on (1) a typical cloud OLAP setting, and (2) a single-node DBMS setting.  For (1), a 4-node Spark cluster is hosted by Azure, each with 8 cores, 56GB RAM, and an SSD. All data is stored on Azure Storage, a remote blob store. For (2), we use an optimized commercial engine.
We report the execution runtime of TPC-H on (1) a distributed Spark cluster, and (2) a single-node commercial DBMS.

% We use a range-partitioning baseline on \textsf{o\_orderdate} and we make sure all layouts contain comparable numbers of blocks.
\vspace{4pt}
\noindent {\bf Distributed Spark.} Figure~\ref{fig:runtime-tpch-spark} shows the mean runtime per template on distributed Spark (each template has 10 random instances).  In total, \qdrt yields a speedup of $1.6\times$ (closely matching the logical ratio in Table~\ref{table:tpch-with-binary}). When excluded templates that need to scan all data, the speedup is $2.6\times$.
% \footnote{We also executed on raw CSV data and the overall trends are similar: the two speedups mentioned above became $1.6\times$ and $3.5\times$, respectively.}.
% Finally,  we see that \qdrt is ``safe'' to use in that it does not slow down queries \zongheng{on second thought, a weak argument?}.

The top three templates where \qdrt exhibits the most absolute runtime reduction are $q_{21}, q_5, q_{19}$. %, q_{17}$.
For $q_{21}$, the advanced cut \textsf{l\_commitdate < l\_receiptdate} allows many blocks to be skipped, substantially speeding up a self-join.
$q_5$ has filters on supplier's \textsf{r\_name}, a categorical with diverse literals---\qdrt yields a $16.8\times$ speedup on this template.
$q_{19}$ is an OR of three complex 6-filter blocks; \qdrt is able to optimize for this complex template and provides $5.5\times$ speedup over Bottom-Up.
Bottom-Up is faster only on $q_1$ and $q_{18}$, both of which require the full month worth of data.
% $q_{17}$ contains a self-join (hence, it needs to touch all data); \qdrt can still skip blocks for one side since it contains selective filters.
% Lastly, $q_{7}$ filters on \textsf{l\_shipdate}, which is correlated with baseline's partitioning column; however, the additional disjunctive filters on two \textsf{n\_name} columns are not handled by baseline Parquet (\qdrt yields a $8.3\times$ speedup on $q_7$).

% {\bf (Single-node OLAP) 1-node SparkSQL.}
% Moving to the setting of single-node OLAP, we run ErrorLog and UserLog workloads on a single-node SparkSQL setup.
% Figure X shows the results.
% The takeaways are YY.

%\todo{no route results? Zongheng: running.}

\vspace{4pt}
\noindent {\bf Commercial DBMS.}
We move to a second execution engine, the commercial DBMS, to investigate whether \qdrt still provides physical runtime execution.
Results are shown in Figure~\ref{fig:runtime-tpch-dbms}. Over all templates, \qdrt has a speedup of $1.3\times$ and when excluding the scan-all templates, the speedup is $1.7\times$.
Consistent with SparkSQL results, templates $q_{21}, q_5, q_{19}$ exhibit large speedups of $1.3\times, 8.6\times, 9.1\times$ respectively.
Relative ratios for other templates are also consistent, which suggests that the benefits from improved layouts carry over. % to  this second engine.
% On this more optimized system, we expect the runtime reduction to be greater than that on SparkSQL, because scan overheads should become more dominant \todo{somehow not true; single-node vs. distributed?}.\\

% \subsubsection{Performance of data and query routing.}
\vspace{4pt}
\noindent {\bf Performance of data and query routing.}
Figure~\ref{fig:data-routing} reports the throughput of routing records through a \qdrt (i.e., ingestion).  We vary the number of ingestion threads on the same machine; linear scalability is achieved with up to 16 threads, and at 64 threads, our prototype implementation in Python (which uses vectorized operations when possible) can reach 400K records/second.  Higher throughput can be reached still, by either an optimized implementation in a compiled language, or by scaling out to multiple nodes.

Figure~\ref{fig:query-routing} shows the latency CDF for routing all 150 queries---the time it takes to check each query against a \qdrt to determine intersecting blocks (Section~\ref{sec:query-proc}).
The maximum time it takes for a query is less than 16ms, with most under 10ms.
The latencies for checking against the semantic descriptions are not homogeneous, because queries have varying number of filters (of varying complexity).

\vspace{4pt}
\noindent {\bf Robustness.} We generated $10\times$ more queries (100 queries per template) using distinct random seeds from before.  This enlarged ``test'' set includes substantially more query literals that the \qdrt did not use for construction.  The mean runtime of these 1500 queries on the same \qdrt layout as Figure~\ref{fig:runtime-tpch-spark} is 7776ms, compared to that figure's 7752ms, the mean of the 150 ``train'' queries.  This suggests that \qdrt is robust for this templated workload with unseen literals.
 }

% In Table~\ref{table:desc-ablation}, we take the best tree produced for the configuration \texttt{min\_size}=50K in Table~\ref{table:tpch}, but evaluate its goodness under different description checks: (1) only per-block min-max index is used, (2) additionally using a categorical column mask (for IN and EQUAL), and (3) further adding a binary cut mask.  The results show that the best quality is realized when all three of \sys's description checks are utilized.  \zongheng{This table may be misleading.  It does not equate \sys producing bad trees when categoricals and binary cuts are excluded in its action space. }

\removedfromrevision{\subsection{Results on Real Workload}}
% \subsection{Results on Real Datasets}
\subsection{ErrorLogs}
\label{sec:errorlog}
We now discuss \revision{ErrorLog-Int and ErrorLog-Ext}. Table~\ref{table:tpch-with-binary} reports the logical metrics for all approaches.
% We now evaluate our techniques on the \revision{ErrorLog-Int and ErrorLog-Ext} datasets in Table~\ref{table:tpch-with-binary}.
% We evaluate the following alternatives: Random partitioning, Ingest partitioning based on client ingest time (the default range partitioning used for this workload), Bottom-Up, greedy \qdrt, and RL \qdrt.

\removedfromrevision{Interestingly, unlike TPC-H, random partitioning performs poorly for \revision{both real
datasets}, with queries having to access all blocks similar to full scan. This is likely due
to the poor filtering provided by such an approach.}
The default range-partitioner (``Baseline'')
accesses all tuples. \revision{Further, we found that the original feature selection method in Bottom-Up ends up choosing a predicate with very high frequency but also very high selectivity as a feature. It prunes other predicates due to the frequency discount used in the original paper, and its skipping power is poor. Thus, the access ratio is close to $100\%$. This a weakness of frequency-based feature selection without taking selectivity into account. We tuned it by ignoring predicates with selectivity > $10\%$. This tuning, which we call $\text{BU}^+$, improves the access ratio of Bottom-Up to $5$--$12\%$ (listed in Table~\ref{table:tpch-with-binary}).} % for our real datasets and is reported in Table~\ref{table:tpch-with-binary}.}

\removedfromrevision{We hypothesize that for this workload, Bottom-Up chooses a dimension weaker than ingestion time as a feature, and that feature prunes out the ingestion time feature, resulting in similar or slightly worse skip quality. }

Our greedy \qdrt, on the other hand, achieves excellent skipping capability, accessing only \revision{$3.1$\% and $1.7$\% of tuples for ErrorLog-Int and ErrorLog-Ext, respectively. The RL \qdrt improves them to $0.4$\% and $0.2$\% respectively, an up to $8.5\times$ improvement}. %or around $7.7$\times$$\% better than our greedy version. \todo{fix numbers, add for Int and Ext separately}.

\begin{figure}[t]
     \centering
     \begin{subfigure}[b]{0.23\textwidth}
         \centering
      \includegraphics[width=\columnwidth]{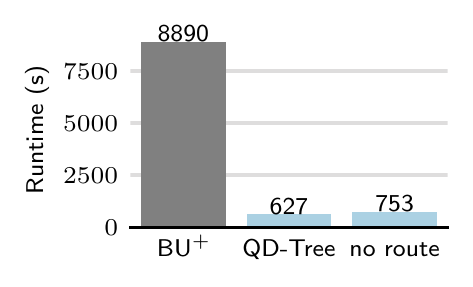}
\vspace{-.33in}
         \caption{\small{ErrorLog-Int}\label{fig:}}
     \end{subfigure}
     \begin{subfigure}[b]{0.23\textwidth}
         \centering
          \includegraphics[width=\columnwidth]{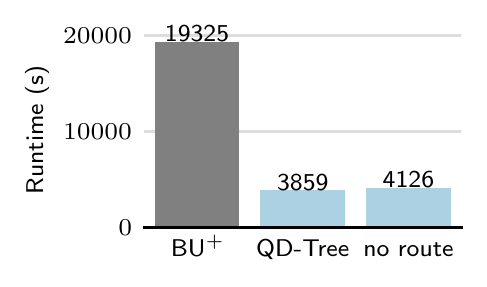}
\vspace{-.33in}
         \caption{\small{ErrorLog-Ext}\label{fig:}}
     \end{subfigure}
     \begin{subfigure}[b]{0.46\textwidth}
         \centering
  \includegraphics[width=\columnwidth]{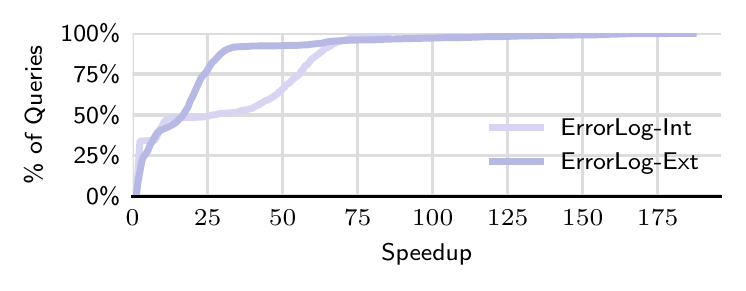}
\vspace{-.33in}
         \caption{\small{CDF of per-query speedups over BU$^+$}\label{fig:}}
     \end{subfigure}
\vspace{-.04in}
\caption{\small \revision{ErrorLog execution runtimes.}
  \vspace{-.15in}
\label{fig:runtime-watson}}
\end{figure}

% 28999997records-380queries-95lits-0.016sel-0.992rew0.984-7epi-237.3imbl-579targetParts-7actual.pkl

% Performing correctness check, which takes a while...

% Tree appears correct.

% all descriptions

% On 28999997 records, 380 total queries: skip ratio 0.9841, access ratio 0.0159

% - binary cuts

% On 28999997 records, 380 total queries: skip ratio 0.9841, access ratio 0.0159

% - categoricals

% On 28999997 records, 380 total queries: skip ratio 0.9841, access ratio 0.0159

% min max only

% On 28999997 records, 380 total queries: skip ratio 0.9841, access ratio 0.0159
% https://github.com/badrishc/DataLayout/blob/row-group/plotdata/real1/summary.log

% \begin{table}[htp]\centering \small%
% \begin{tabular}{@{} l l l l l @{}} \toprule
%   &  Ingest  & Bottom-Up & Greedy \qdrt & RL \qdrt  \\ \midrule
%   \revision{Int \%} & \todo{}\% & 81.3\% & 2.09\% & 1.59\% \\
%   \revision{Ext \%} & \todo{}\% & 81.3\% & 2.09\% & 1.59\% \\
% \bottomrule
% \end{tabular}
% \caption{\small{{\revision{ErrorLog-Int and ErrorLog-Ext}: Percentage of tuples accessed under different layout schemes, compared to full scan. The workload consists of \revision{1000} queries. Random gives no skippability, and is not shown. \todo{This table needs updates?}} }\label{table:corp}}
% \end{table}

\revision{

\subsubsection{Physical execution}
\label{sec:watson-physical}

We measure the actual execution runtime for both real datasets on an optimized single-node SparkSQL instance.
% We first create Parquet versions of the datasets based on Ingest, Bottom-Up$^+$, and \qdrt based partitioning techniques, making sure all layouts contain a comparable number of blocks.
% We execute each of the $\sim 1000$ queries in each workload over the two datasets.
Each workload consists of $1000$ queries.
Figures~\ref{fig:runtime-watson}(a) and (b) show the aggregate runtimes.

% We do not depict Ingest as it needs to perform a full scan.
We find that \qdrt dominates Bottom-Up$^+$ with a $14\times$ lower runtime for ErrorLog-Int. On ErrorLog-Ext, the speedup becomes $5\times$ because of its higher selectivity. For \qdrt, we report runtimes using \qdrt routing (adding {\tt BID IN (...)} predicates) and using the default partition pruning (\emph{no route}). On SparkSQL with Parquet, we observe that \qdrt-based routing is better than \emph{no route} by $16$\% for ErrorLog-Int and $6.4$\% for ErrorLog-Ext.

Figure~\ref{fig:runtime-watson}(c) shows the CDF of per-query speedups. We observe that $50\%$ of queries have a speedup of at least $25\times$ (ErrorLog-Int) and $20\times$ (ErrorLog-Ext), respectively.  Thus, the excellent skipping benefits reflected in the logical metrics translate well into physical runtime reduction.
% more than $\todo{}$\% of queries have a speedup of at least $10\times$, while some queries getting more than $\todo{rephrase}\times$ speedups.

Lastly, we executed a subset of the ErrorLog-Int queries on the commercial DBMS. Briefly, the trends remain consistent as before. Query execution with \qdrt query routing is $\sim 400\times$ faster than the range-partitioned baseline which needs to touch all blocks. Unlike Parquet, however, we found that \emph{no route} performed significantly worse than \qdrt routing. We believe this is due to a lack of block-level indexes (dictionaries) for categorical fields, which prevents categorical predicates from pruning blocks in \emph{no route}.
}

\subsection{Time to Produce Layouts}
\label{sec:time-to-layout}
While quality of data layout is the primary metric of concern in this study, we next discuss the wall-clock time required to produce good data layouts.

\begin{figure}[t!]
\centering
\includegraphics[width=\columnwidth]{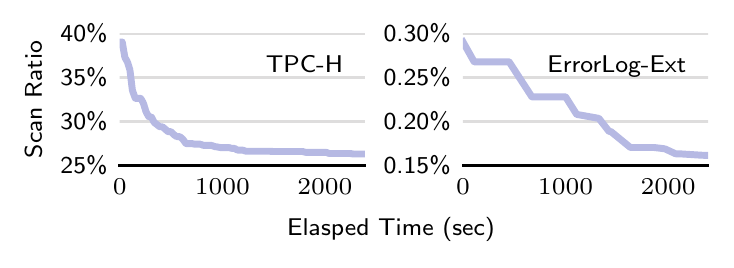}
\vspace{-.3in}
\caption{\small Learning curve of \sys.  On TPC-H, most quality improvement is learned in the first $\sim$10 minutes; \revision{on ErrorLog-Ext, high quality is achieved immediately ($.3\%$) and continuously improved when given more time budget. The trend for ErrorLog-Int is similar (not shown).}
\vspace{-.20in}
% We show non-learning baselines markers as comparisons. 
\label{fig:learning-curve}}
\end{figure}

\textbf{TPC-H}. Bottom-Up took %$4356$ secs
\revision{71 minutes} to produce its layout due to large number of records and queries.
% , whereas the Greedy \qdrt construction took
% a shorter time of $1246$ secs \todo{update}
% \revision{370 minutes}
% without any optimizations.
% Both Bottom-Up and Greedy produce a layout on termination.
It produces a layout only on termination.
In contrast, the RL agent \sys produces trees immediately and continuously, with quality improving over time. \removedfromrevision{Hence we discuss this technique in more detail next.}
Figure~\ref{fig:learning-curve} plots the learning curves of \sys. There are several key takeaways.

First, at random initialization, \sys immediately produces partitioning trees
with a scan ratio of \revision{$\sim 39\%$}. % $\sim 12\%$.
This, in fact, significantly exceeds the quality of the
Random partitioner reported in Table~\ref{table:tpch-with-binary} \revision{($56\%$)}. %($28.8\%$).
The reason is that, at initialization \sys produces a tree \emph{drawn randomly from the search space}, which is defined with a candidate cut set extracted from query workload.
Leveraging these informative cuts is much better than disregarding workload information.

% \vspace{-.1in}

Second, \sys enables the user to explicitly trade-off computation time vs. the quality of layouts produced.
\removedfromrevision{It is the only approach in our setup where the user could spend more machine time, if the budget allows, to obtain continuously improved layouts (up to convergence).}
As the agent constructs more trees, it learns to bias the cuts that are observed to more profitable, but it also keeps a non-zero probability for exploration.
\eat{Conversely, if time to produce layout is under a stringent deadline, the user is free to take lesser-quality trees.}
We see that most of the improvement is learned in the first 10 minutes.
Further speedups may result from (1) implementing our tree library in a native language rather than in Python, and (2) switching \sys's learning algorithm to
a distributed learner~\cite{impala}.%
% those designed to be run at large scale with high throughput~\cite{impala}.

\textbf{ErrorLogs.} A uniform workload like TPC-H is, in fact, a more challenging task for RL because the uniform distribution of data and queries has higher entropy. On both real ErrorLog workloads, \sys produces top-performing trees within 30 seconds (Fig.~\ref{fig:learning-curve}).  The run time is much shortened, due to the abundant correlations to be exploited in the real-world data/workload. With the existence of correlations, exploration of the space is significantly faster.
\revision{
On ErrorLog-Int, Greedy and Bottom-Up finished in 12 minutes and 432 minutes, respectively.
On ErrorLog-Ext, these numbers became 12 minutes and 565 minutes, respectively.
}

\subsection{Interpreting Learned \qdrts}

To gain insights, we now analyze a top-performing \qdrt (constructed by \sys) found for the TPC-H workload.

\begin{figure}[t!]
\centering
\includegraphics[width=\columnwidth]{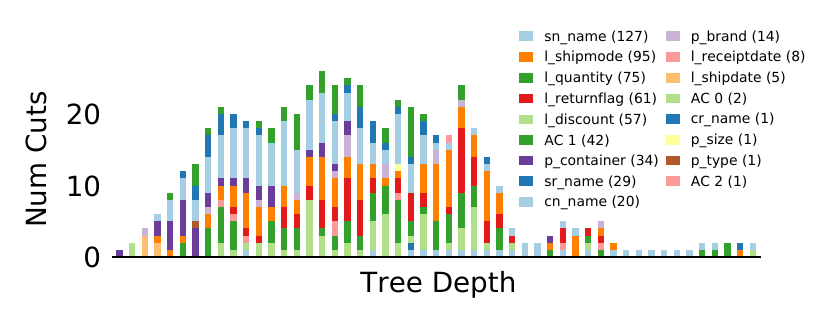}
\vspace{-.25in}
\caption{\small A \sys-produced top-performing \qdrt for TPC-H.  The number after each legend indicates the total number of cuts on that column (or advanced cut).\label{fig:tpch-tree}}
\vspace{-.20in}
\end{figure}

Figure~\ref{fig:tpch-tree} plots the dimensions cut across tree levels.  We make a few observations.  First, \emph{the variety of cuts is high}: 8 columns are cut at least 20 times throughout all tree levels.  Both categorical (e.g., \texttt{l\_shipmode}) and numerical columns (e.g., \texttt{l\_discount}) contribute to skipping.  %Second, such
This indicates fine-grained cutting, as is done by RL-learned \qdrt, is beneficial.
Second, \emph{advanced cuts are leveraged} (AC; Sec.~\ref{sec:adv-cuts}), indicating their skippability benefits in complex workloads.
% Second, \sys is able to find such fine-grained cutting, indicating its potential advantage in navigating more complex workloads.

% We also report the first three cuts made at the root and at the first two children:
The cuts made at the root and the first two children are:
\begin{itemize}[leftmargin=*]
\item \texttt{p\_container IN (LG CASE,LG BOX,LG PACK,LG PKG)};
\item (First two children) $AC_0$: \texttt{c\_nationkey = s\_nationkey}.
\end{itemize}
% (1) \texttt{l\_shipmode='AIR'}, (2) \texttt{l\_shipinstruct='DELIVER IN PERSON'}, and lastly (3) \texttt{r\_name='MIDDLE EAST'};
The next level's cuts involve \texttt{l\_shipdate} and \texttt{p\_brand}.
Overall, it is clear that existing partitioning techniques (hash or range) do not equate the sophisticated combination of cuts produced by a \qdrt layout.
% Note that the second cut is the only cut on \texttt{l\_shipinstruct}, which implies naively partitioning by that column does not equate to this \qdrt's partitioning (it is defined by a complex combination of cuts across all levels).

% LocalWords:  ErrorLog shipmode shipinstruct Greedy's ErrorLogs nationkey
% LocalWords:  shipdate

\section{Related Work}
\label{sec:related}

\revision{
\noindent
{\bf Physical Design \& Partitioning.} Traditionally, data warehouses employ partitioning for the purpose of scaling out computation and load balancing. Data is either chunked into blocks based on arrival time, or partitioned using range or hash partitioning schemes~\cite{DBLP:conf/sigmod/LarsonCFHMNPPRRS13} or their improvements~\cite{Bhattacharjee2003EfficientQP, Zhou:2012:APT:2213836.2213839}. Our technique may be applied within partitions created by such static schemes. Automated physical design tools provide auto-tuning capabilities based on what-if analyses and integrated into the query optimizer~\cite{Agrawal,agrawal2006autoadmin,Rao:2002:APD:564691.564757, 10.1145/42201.42205, harder1976selecting}. AutoAdmin~\cite{Agrawal,agrawal2006autoadmin, zilio2004db2, olma2017slalom, pavlo2012skew, wu2011partitioning} optimizes a database and its physical design using machine learning and data mining. Some systems perform partitioning based on workload access patterns~\cite{Agrawal, arulraj2016bridging, chasseur2013design}, while other systems are based on graph-based workload modeling techniques~\cite{curino2010schism, quamar2013sword, serafini2016clay}. Sun et al.~\cite{bottom_up, DBLP:journals/pvldb/SunFWW16} (discussed in this paper) extract features from each workload operation based on its predicates. Casper~\cite{athanassoulis2019optimal} offers a general partitioning design tool based on navigating a three-way tradeoff between read performance, update performance, and memory utilization.}

\revision{
  \vspace{5pt}
\noindent
{\bf Adaptive Physical Design.}
The cost of periodic automated physical design can be amortized in an online pay-as-you-go fashion. This line of research can roughly be categorized into two approaches. Online analysis~\cite{bruno2006tune, bruno2007online, schnaitter2006colt} uses available system cycles between query execution and offline analysis to optimize physical design. Database cracking~\cite{idreos2007database, idreos2011merging} immediately starts executing queries, and treats each query as a hint to reorganize parts of the data during query processing, usually based on sort attributes. This approach optimizes and adapts the data layout over time.

In contrast, a \qdrt performs data layout using fine-grained descriptions based on a workload set. Our technique learns at coarse-grained time boundaries and emits a synthesized \qdrt data structure to enable data layout. We do not react to individual query execution, but our resulting data structure may be used for continuous ingestion or bulk loading. We find this approach to be suitable for modern data formats such as Parquet, where incremental re-organization is expensive. An interesting direction of future work is to integrate cracking with \qdrt. Since a \qdrt represents a way to layout data, cracking would allow us to incrementally refine the \qdrt over time. A possible approach, left as future work, would be to include the cost of data re-organization into the \qdrt cost model, so that we can enable efficient incremental re-organization over time.}

\revision{
  \vspace{5pt}
\noindent
{\bf Learned Databases.}
There is increasing interest in automating core database functionality and design decisions.
This line of research leverages recent advancements in deep learning algorithms and scalable hardware (GPU) to improve database systems.
Closest to our work is a proposal of learned partition adviser using deep RL~\cite{hilprecht2019towards};
% it focuses on traditional (e.g., hash) partitioning and replication and does not consider query predicates.
it focuses on replication and coarse-grained partitioning (e.g., hash) along entire attribute(s), unlike \qdrt which partitions based on a rich set of fine-grained candidate cuts.
In this space, machine learning has also been used to revisit tuning~\cite{van2017automatic}, workload forecasting~\cite{ma2018query}, data structures and indexes~\cite{kraska2018case, idreos2019design, nathan2019learning,ding2019alex}, and query optimization~\cite{naru,dq,neo,dutt2019selectivity}.
% Recent work revisits physical design in this light~\cite{van2017automatic, dayan2017monkey, idreos2019design, dayan2018optimal, olma2017slalom}.
% Also in this space is the notion of workload forecasting~\cite{ma2018query} using machine learning, which further supports workload-optimized database research.
% Further, machine learning for systems has seen increased attention, with ML applied to data structures and indexes~\cite{kraska2018case, dayan2017monkey, dayan2018optimal, idreos2019design} and query optimization~\cite{naru,dq,dutt2019selectivity}.
Our \qdrt may be viewed as a learned physical design or indexing tool: it optimizes for scan-based workloads, common in big data analytics, to minimize the I/O cost of block accesses.
% is workload-optimized for the specific purpose of block creation and access in scan-based analytical workloads with a view to minimizing the I/O cost of block access.
}

\removedfromrevision{
\noindent
{\bf Learned Decision Trees via Deep RL.}  Our \sys design is inspired by
NeuroCuts~\cite{neurocuts}, a deep RL algorithm to design decision trees for
network packet classification.  As already described in detail in
Section~\ref{sec:rl-related}, we adopt their tree-structured MDP formulation, and
the rest of \sys design (state space, action space, rewards, optimizations) is
specific to our SQL and data layout settings.
}

\noindent
{\bf Traditional Indexing.} Database indexing is a well-studied space. For one dimensional indexing, B-Trees form the state of the art. Multi-dimensional indexes such as \emph{k}-d tree~\cite{bentley1975multidimensional} and R-tree~\cite{guttman1984r} have been proposed to index data over more than one dimension, but they do not adapt to high dimensional data or the specific query workload. These indexes are not a good fit for analytical workloads due to the cost of each index lookup. Modern systems instead use scan-oriented processing over columnar row-groups. \revision{Our work may be viewed as workload-aware multi-dimensional indexing adapted to the needs of analytical scan-based workloads.}

\revision{
  \vspace{5pt}
\noindent
{\bf Partition Pruning.} Most scan-oriented databases employ indexing over the blocks to make it easy to skip blocks. Examples include min-max based pruning, also known as small materialized aggregates (SMA)~\cite{moerkotte1998small}, zone maps~\cite{graefe2009fast}, and data skipping~\cite{bottom_up, DBLP:journals/pvldb/SunFWW16}. Here, the system maintains the data distribution information for each block. Depending on the query
predicates, these values can be used to determine that a given block  might not be needed for a given query.} They are commonly employed in systems such as Oracle~\cite{oracle}, Postgres~\cite{postgres}, Microsoft SQL Server~\cite{DBLP:conf/sigmod/LarsonCFHMNPPRRS13}, \revision{and Snowflake~\cite{dageville2016snowflake}}. The most popular SMA is the min-max index, which maintains the minimum and maximum value for each field, per block. \revision{Snowflake also maintains SMAs for auto-detected columns in semi-structured data. They are used for simple predicates as well as more complex predicates, such as {\tt IN}.} We leverage such block-level indexes in this paper as well. During query processing, we use a combination of tree routing and SMA indexes to achieve maximum skippability.

\removedfromrevision{Other aggregates such as sum, count, and histograms may also be materialized per block for various fields -- these techniques are orthogonal to our contributions and may be leveraged in addition to our data structures.}

% Recently, machine learning for systems has seen increased attention, with ML applied to operations such as indexing~\cite{kraska2018case} and selectivity estimation~\cite{dutt2019selectivity}. 

\removedfromrevision{
\noindent
{\bf View Selection.} Our work is also related to the problem of selecting materialized views over data, such as the recent work on CloudViews~\cite{Jindal:2018:SSM:3192965.3228338}. However, our layout techniques apply at the storage level and not at the query processing level. Said differently, our storage techniques may be used at a level below standard schemes used for materialized view selection that may exploit complex query semantics to determine what views to store and maintain.
}

\removedfromrevision{That said, our semantically complete blocks may be seen as a form of materialized views over the primary copy of the data, because we support semantic block descriptions with the completeness property. Finally, materialized view selection often optimizes for the view maintenance cost~\cite{gupta1999selection}, which we do not consider for data layout.}

% LocalWords:  Bw AutoAdmin CloudViews NeuroCuts

\section{Conclusion}
% \section{Conclusions}
\label{sec:conclude}

% The need to run queries at interactive speeds on large datasets is increasingly important.
Running queries at interactive speeds on large datasets is increasingly important.
In this paper, we address the problem of best assigning records to data blocks on storage, with a goal to optimize for the important metric of \emph{number of blocks accessed} by a query. This metric directly relates to the I/O cost, and therefore performance, of most analytical queries. Current techniques based on hash and time-based partitioning or clustering are unable to exploit the workload fully and provide blocks with complete semantic descriptions, which is useful for caching, exploiting additional storage, etc. We propose a new framework called a \emph{query-data routing tree}, or {\em\qdrt}, to address this gap. Further, we develop two novel algorithms for \qdrt construction: one based on a greedy approach and the other based on deep reinforcement learning. Experiments over benchmark and real workloads show that a \qdrt can provide physical speedups of more than an order of magnitude compared to current blocking schemes, and can reach within $2\times$ of the lower bound for data skipping based on selectivity, while providing complete semantic descriptions of created blocks.

% Experiments over benchmarks and real workloads indicate that we are able to outperform the quality of current blocking schemes by up to $14\times$, while providing complete semantic descriptions of created blocks.
% The need to run queries at interactive speeds on very large datasets is becoming increasingly important. In this paper, we address the problem of best assigning records to data blocks on storage, with a goal to optimize for the important metric of \emph{number of blocks accessed} by a query. This metric directly relates to the I/O cost, and therefore performance, of most analytical queries. Current techniques based on hash and time-based partitioning or clustering are unable to exploit the workload fully and provide blocks with complete semantic descriptions, which is useful for caching, exploiting additional storage, etc. We propose a new framework called a \emph{query-data routing tree}, or {\em\qdrt}, to address this gap. Further, we develop two novel algorithms for \qdrt construction: one based on a greedy approach and the other based on deep reinforcement learning. Experiments over benchmarks and real workloads indicate that we are able to outperform the quality of current blocking schemes by up to an order-of-magnitude, and reach within $2\times$ of the lower bound for data skipping based on selectivity, while providing complete semantic descriptions of created blocks.

\balance
\bibliographystyle{ACM-Reference-Format}
\bibliography{dl-sigmod20}

\end{document}